\documentclass[structabstract,longautha]{aa}
\usepackage{natbib}
\usepackage{graphicx}
\usepackage{txfonts}
\newcommand{\NH}{N$_{\mbox{\scriptsize{H}}}$}
\newcommand{\Halpha}{\ifmmode {\rm H}\alpha \else H$\alpha$\fi}
\newcommand{\Hbeta}{\ifmmode {\rm H}\beta \else H$\beta$\fi}
\newcommand{\Hgamma}{\ifmmode {\rm H}\gamma \else H$\gamma$\fi}
\newcommand{\Hdelta}{\ifmmode {\rm H}\delta \else H$\delta$\fi}
\newcommand{\Lya}{\ifmmode {\rm Ly}\alpha \else Ly$\alpha$\fi}
\newcommand{\Lyb}{\ifmmode {\rm Ly}\beta \else Ly$\beta$\fi}
\newcommand{\HeI}{\ifmmode {\rm He}\,\textsc{i}\,\lambda5876 \else 
	          He\,\textsc{i}\,$\lambda5876$\fi}
\newcommand{\HeII}{\ifmmode {\rm He}\,\textsc{ii}\,\lambda4686 \else 
	           He\,\textsc{ii}\,$\lambda4686$\fi}

\newcommand{\neiii}{[Ne\,\textsc{iii}]}

\newcommand{\nev}{[Ne\,\textsc{v}]}

\newcommand{\ciii}{\ifmmode {\rm C}\,\textsc{iii} \else C\,\textsc{iii}\fi}
\newcommand{\civ}{\ifmmode {\rm C}\,\textsc{iv} \else C\,\textsc{iv}\fi}

\newcommand{\nii}{[N\,\textsc{ii}]}

\newcommand{\oii}{[O\,\textsc{ii}]}
\newcommand{\oiii}{[O\,\textsc{iii}]}

\newcommand{\mgii}{Mg\,\textsc{ii}}

\newcommand{\tyi}{\hbox{Type-1}}
\newcommand{\tyii}{\hbox{Type-2}}
\begin{document}
%
   \title{Obscured AGN at $z\sim\,$1 from the zCOSMOS-Bright Survey}
   \subtitle{I. Selection and Optical Properties of a \nev-selected sample}
\author{
M.~Mignoli\inst{1}
\and
C.~Vignali\inst{2}
\and
R.~Gilli\inst{1}
\and
 A.~Comastri\inst{1}
\and
G.~Zamorani\inst{1}
\and
M.~Bolzonella\inst{1}
\and
A.~Bongiorno\inst{3}
\and
F.~Lamareille\inst{4,5}
\and
P.~Nair\inst{1,6}
\and
L.~Pozzetti\inst{1}
\and
S.J.~Lilly\inst{7}
\and
C.M.~Carollo\inst{7}
\and
T.~Contini\inst{4,5}
\and
J.-P.~Kneib\inst{8}
\and
O.~Le~F\'evre\inst{8}
\and
V.~Mainieri\inst{9}
\and
A.~Renzini\inst{10}
\and 
M.~Scodeggio\inst{11}
\and
S.~Bardelli\inst{1}
\and
K.~Caputi\inst{12}
\and
O.~Cucciati\inst{13}
\and
S.~de~la~Torre\inst{14}
\and
L.~de~Ravel\inst{14}
\and
P.~Franzetti\inst{11}
\and
B.~Garilli\inst{11}
\and
A.~Iovino\inst{13}
\and
P.~Kampczyk\inst{7}
\and
C.~Knobel\inst{7}
\and
K.~Kova\v{c}\inst{7}
\and
J.-F.~Le~Borgne\inst{4,5}
\and
V.~Le~Brun\inst{8}
\and
C.~Maier\inst{7,15}
\and
R.~Pell\`o\inst{4,5}
\and
Y.~Peng\inst{6}
\and
E.~Perez~Montero\inst{4,5}
\and
V.~Presotto\inst{13}
\and
J.D.~Silverman\inst{16}
\and
M.~Tanaka\inst{16}
\and
L.~Tasca\inst{8}
\and
L.~Tresse\inst{8}
\and
D.~Vergani\inst{17}
\and
E.~Zucca\inst{1}
\and
R.~Bordoloi\inst{7}
\and
A.~Cappi\inst{1}
\and
A.~Cimatti\inst{2}
\and
A.M.~Koekemoer\inst{6}
\and
H.J.~McCracken\inst{18}
\and
M.~Moresco\inst{2}
\and
N.~Welikala\inst{19}
}

\institute{
  INAF -- Osservatorio Astronomico di Bologna, via Ranzani 1, 40127 Bologna, Italy\\
    \email{marco.mignoli@oabo.inaf.it}
\and
  Dipartimento di Fisica e Astronomia, Universit\`a degli Studi di Bologna, viale Berti Pichat 6/2, 40127 Bologna, Italy
\and
  INAF -- Osservatorio Astronomico di Roma, 00040, Monteporzio Catone, Italy
\and
  Institut de Recherche en Astrophysique et Plan\'etologie, CNRS, F-31400 Toulouse, France
\and
  IRAP, Universit\'e de Toulouse, UPS-OMP, Toulouse, France
\and
  Space Telescope Science Institute, Baltimore, MD 21218, USA
\and
  Institute of Astronomy, ETH Zurich, 8093 Z\"urich, Switzerland
\and
  Laboratoire d'Astrophysique de Marseille, Aix Marseille Universit\'e, CNRS, Marseille, France
\and
  European Southern Observatory, Garching, Germany
\and
  INAF -- Osservatorio Astronomico di Padova, Padova, Italy
\and
  INAF -- Istituto di Astrofisica Spaziale e Fisica Cosmica, Milano, Italy
\and
 Kapteyn Astronomical Institute, University of Groningen,  9700 AV Groningen, The Netherlands
\and
  INAF -- Osservatorio Astronomico di Brera, Milano, Italy
\and
  Institute for Astronomy, The University of Edinburgh, Royal Observatory, Edinburgh, EH93HJ, UK
\and
University of Vienna, Department of Astronomy, 1180 Vienna, Austria
\and
Kavli Institute for the Physics and Mathematics of the Universe, The University of Tokyo, Kashiwa 277-8583, Japan
\and
  INAF -- Istituto di Astrofisica Spaziale e Fisica Cosmica, Bologna, Italy
\and
Institut d'Astrophysique de Paris, Universit\'e Pierre \& Marie Curie, 75014 Paris, France
\and
Institut d'Astrophysique Spatiale, Batiment 121, CNRS \& Univ. Paris Sud XI, 91405 Orsay Cedex, France
}

   \date{Received ; accepted }

 
  \abstract
   {}
   {The application of multi-wavelength selection techniques is essential for
   obtaining a complete and unbiased census of active galactic nuclei (AGN).
   We present here a method to select $z\sim\,$1 obscured AGN from
   optical spectroscopic surveys.}
   {A sample of 94 narrow~line AGN with 0.65$\,<\,${\it z}$\,<\,$1.20 has been selected
    from the 20k-Bright zCOSMOS galaxy sample by detection of the high-ionization
    \nev~$\lambda3426$ line. The presence of such emission line in a galaxy
    spectrum is indicative of nuclear activity, although the selection is biased toward
    low absorbing column densities on narrow line region or galactic scale.
    A~similar sample of unobscured (\tyi~AGN) has been collected applying the same analysis to
    zCOSMOS broad-line objects. 
    This paper presents and compares the optical spectral properties of the two AGN samples.
    Taking advantage of the large amount of data available
    in the COSMOS field, the properties of the \nev-selected \tyii~AGN have been
    investigated, focusing on their host galaxies, X-ray emission, and optical line flux ratios.
    Finally, the diagnostic developed by Gilli et~al. (2010), based on the X-ray to \nev\
    luminosity ratio, has been exploited to search for the more heavily obscured AGN.}
   {We found that \nev-selected narrow line AGN have Seyfert~2-like optical spectra,
    although with emission line ratios diluted by a star-forming component. 
    The ACS morphologies and stellar component in the optical
    spectra indicate a preference for our \tyii~AGN to be hosted in early-spirals
    with stellar masses greater than $10^{9.5-10}M_\odot$,
    on average higher than those of the galaxy parent sample.
    The fraction of galaxies hosting \nev-selected obscured AGN increases
    with the stellar mass, reaching a maximum of about 3\% at
    $\approx$2$\times\,10^{11}M_\odot$.
    A comparison with other selection techniques at {\textsl z}$\sim$1,
    namely the line-ratio diagnostics and X-ray detections, shows that the detection
    of the \nev~$\lambda3426$ line is an effective method to select AGN
    in the optical band, in particular the most heavily obscured ones, but can
    not provide by itself  a complete census of \tyii~AGN. Finally, the high fraction
    of \nev-selected \tyii~AGN not detected in medium-deep ($\approx$100-200~ks) 
    Chandra observations (67\%) is suggestive of the inclusion of Compton-thick
    (i.e. with \NH$>10^{24}cm^{-2}$) sources in our sample. 
    The presence of a population of heavily obscured AGN is corroborated
    by the X-ray~to~\nev~ratio;  we estimated, by mean of X-ray stacking
    technique and simulations, that the Compton-thick fraction in our sample of
    \tyii~AGN is $43\pm4\%$ (statistical errors only), in good agreement with  
    standard assumptions by the XRB synthesis models.
    }
   {}


   \maketitle
%

\section{Introduction}
   The study of the history of accretion is essential to our
   understanding of how supermassive black holes (SMBHs) form and evolve.
   Accretion onto a SMBH is the predominant source of energy emitted
   by Active Galactic Nuclei (AGN), so a comprehensive census of AGN
   of all types across a large fraction of cosmic time provides
   constraint on the black hole mass function at the present day
   \citep{Soltan82, Rees84, Marconi04}.
   
   Nowadays, there are strong observational evidences that all massive
   galaxies in the local Universe host a central SMBH \citep{KormRich95}.
   This, along with the now firmly established discovery that the masses
   of SMBHs are proportional to the velocity dispersions and masses of
   their host stellar spheroids \citep{Mago98, FerMer00, Gebh00, Trema02}, 
   indicates an enduring physical connection between nuclear activity and
   galaxy formation and evolution. Many theoretical and observational
   efforts have been recently undertaken to comprehend the elusive
   evolutionary connection between AGN and their host galaxies. 
   Nevertheless, the mechanisms driving this co-evolution are still
   far from being fully understood. Once again, to clarify the role
   played by AGN in this symbiosis requires a complete survey of both
   unobscured and obscured AGN.

   Active Galactic Nuclei present a large variety of observed properties.
   They inhabit host galaxies of different morphologies,
   and show a wide range of luminosities in all the wavebands, from radio to X-rays.
   In the optical/UV range, AGN are characterized by a power-law
   continuum and broad ($>$1000 km~s$^{-1}$) emission lines, the
   latter produced in the so-called broad-line region (BLR), extending on
   scales of light days. These unobscured (\tyi) AGN are thus easily
   identified from their optical spectra.
   Conversely, there are obscured (\tyii) AGN that show only narrow 
   ($<$1000 km~s$^{-1}$) emission lines emerging from the narrow-line
   region (NLR), with scales of the order of hundreds of light years.
   Since in these AGN the continuum is often dominated by stellar emission,
   their optical spectra are similar to those of normal Star-Forming Galaxies (SFGs).
   According to the standard unified model \citep{Anto93}, 
   this broad classification into two spectral classes depends on
   whether the central SMBH, its associated continuum, and the BLR are
   viewed directly (\tyi~AGN) or are obscured by a dusty circumnuclear
   medium (\tyii~AGN).

   The optical classification of emission-line galaxies is usually
   done through emission-line ratio diagnostic diagrams. \citet{BPT81}
   were the first to apply such a technique (BPT), using the strongest
   emission lines to separate objects into categories according to the
   excitation mechanism of the emitting gas. In particular, the
   \oiii/\Hbeta \ versus \nii/\Halpha \ diagram became the benchmark
   for emission-line classification, since it can reliably
   distinguish Star-Forming galaxies, Seyfert~2 galaxies, Low Ionization
   Nuclear Emission Regions \citep[hereafter LINERs, see][]{Heckman80},
   and composite objects with both an AGN and star-forming regions.
   At redshifts greater than z$\approx$0.5, the emission lines around
   \Halpha \ get redshifted out of the optical window and 
   classical BPT diagrams can no longer be applied. Therefore, diagnostic
   diagrams need to be based on emission lines observed in the blue part
   of the galaxy spectrum: \oiii, \Hbeta, and \oii \ \citep{Rola97}.
   Unfortunately, such diagrams are only moderately effective in discriminating
   between starbursts and AGN \citep{Stasinska06} and
   in particular almost fail to separate LINERS from SFGs \citep{Bongiorno10,
   Lamare10}.

   The X-ray emission is probably the most prominent characteristic of AGN
   activity, since X-rays are thought to originate from the innermost regions
   of an accretion disk around the central SMBH. Combined with observations
   at different wavelengths, deep and wide-field X-ray surveys have been
   indeed effective in discovering a large proportion of the AGN population,
   significantly improving our census of AGN demographics
   \citep[][and references therein]{BH05}. However, a large population of heavily
   obscured AGN, predicted by synthesis models for the cosmic X-ray Background
   \citep[XRB;][]{GCH07}, is still undetected in X-ray surveys.
   Although X-ray observations are the least biased against moderately obscured AGN,
   still even the deepest X-ray surveys can under-sample the population
   of extremely obscured AGN, i.e. those with column densities 
   \NH$>$$10^{24}cm^{-2}$ (i.e. Compton-thick sources).
   
   The presence of the emission line \nev$\,\lambda$3426 can be considered
   a reliable signature for nuclear activity, given that the ionization potential
   of Ne4+ is $\approx$97~eV, and stars generally do not emit
   photons beyond 55~eV \citep{Haehnelt01}. Along with
   the \nev$\,\lambda$3426 line, also the \nev$\,\lambda$3346 line
   arises from the same excitation level, but with a relative intensity of
   approximately one-third \citep{VB_SDSS01}. 
   A \nev-selected AGN sample should not be affected by significant
   contamination from star-forming galaxies: in their large spectroscopic sample,
   \citet{BPT81}  found no [H~\textsc{ii}]~region with detectable \nev,
   although some rare exceptions are known today \citep{Izotov04}.
   Therefore, \nev \ provides a powerful diagnostic tool in detecting
   AGN \citep{Schmidt98,Gilli10}.

   In this paper we report on the first statistical optical \hbox{sample}
   (a total of 94 objects included in our \tyii~AGN class) of \nev-selected
   AGN found in spectra taken in the zCOSMOS Survey \citep{Lilly07}.
   Throughout the paper, we adopt a ``concordance" cosmology, 
   $\Omega_M\,$=$\,0.25$, $\Omega_\Lambda\,$=$\,0.75$,
   and $h\,$=$\,0.7$. Magnitudes are given in the AB system.


\section{The zCOSMOS-Bright spectroscopic survey}

   The Cosmic Evolution Survey \citep[COSMOS,][]{Scoville07} provided
   superb angular resolution and depth with single-orbit I-band
   HST-ACS exposures over a 2 square degree equatorial field \citep{Koeke07},
   along with deep ground-based images with excellent seeing \citep{Capak07}.
   The zCOSMOS spectroscopic survey \citep{Lilly07} yields
   spectroscopic redshifts for a large numbers of galaxies in the 
   COSMOS field using VIMOS, a multi-slit spectrograph mounted
   on the 8m UT3 of the European Southern Observatory's Very
   large Telescope (ESO VLT). The
   zCOSMOS redshift project has been designed, in order to efficiently
   utilize VIMOS, by splitting the survey into two parts:
   first, the zCOSMOS-bright is a pure-magnitude limited survey, which
   has spectroscopically observed about twenty thousands objects
   brighter than \hbox{$I\,$=$\,22.5$} across the entire COSMOS field, with a 
   medium-resolution grism (R$\sim$600) and over a red spectral
   range (5500--9700~\AA). This selection culls galaxies mainly in
   the redshift range \hbox{0.1$\,<\,${\it z}$\,<\,$1.2.}
   The second part, zCOSMOS-deep, has targeted about 10,000 \hbox{$B$$\,<\,$25} 
   galaxies, selected within the central 1 deg$^2$, using color-selection
   criteria to encompass the redshift range \hbox{1.4$\,<\,${\it z}$\,<\,$3.0.}
   In this case, observations were performed with the \hbox{R$\sim$200} LR-Blue
   grism, which provides a spectral coverage from 3600 to 6800~\AA.
   Because of the spectral and redshift ranges covered, the vast majority
   of the zCOSMOS-deep spectra sample a rest-frame wavelength interval
   shortward of 3000\AA. Thus, in this paper we used only the zCOSMOS-bright
   spectra in the search for the \nev$\,\lambda$3426 emission line.
   The spectra were reduced and calibrated using the VIMOS
   Interactive Pipeline Graphical Interface software \cite[VIPGI,][]{Scodeggio05}, 
   while the redshift measurements were obtained with the help of
   an automatic package \citep[EZ,][]{Garilli10} and then visually
   checked. For more details about the zCOSMOS survey,
   we refer the reader to \citet{Lilly09}.  


\section{The \nev-selected Samples}

 \subsection{Samples Selection}

   The 20k zCOSMOS-bright sample (hereafter the zCOSMOS sample) has
   been built from a magnitude-limited survey, with a fairly high sampling rate
   ($\approx$70\%) and well understood completeness properties
   \citep{Zucca09}, that makes it well suited for statistical studies of AGN.
   Among the 20707 entries included in the zCOSMOS spectroscopic catalog
   (version 4.2), a total of 18141 galaxies and AGN have measured redshifts,
   1031 objects (5\%) are spectroscopically classified as stars, and 1535
   objects (7\%) remain without redshift identification. Of course, the success
   rate in securing a redshift is a function of redshift itself, rest-frame color
   and magnitude, but \citet{Lilly09} showed that it is very high ($\approx\,$95\%)
   for the zCOSMOS sample between \hbox{0.5$\,<\,${\it z}$\,<\,$0.9.} 
   In order to guarantee that both the \nev$\,\lambda\lambda$3346,3426
   emission lines fall within our spectral coverage, we have  
   limited our analysis to redshifts larger than 0.65. 
   The standard zCOSMOS redshift measurement process includes several
   interactive steps in which broad line AGN were identified and flagged; 
   initially, we excluded from analysis the already identified \tyi \ AGN,
   because our principal aim is to collect a sample of obscured AGN.
   The analyzed sample contains 7624 galaxies with redshifts between 0.65 and 
   1.50, and the corresponding 8188 spectra\footnote{In the 20k zCOSMOS sample,
   for various reasons more than 5\% of the targets were observed twice or more
   \citep[see][]{Lilly09}.}  were then visually inspected and measured to identify the
   \nev-emitting objects.
   
   The careful \nev-selection procedure consisted of two main steps. First, all the
   galaxy spectra were smoothed with a 3-pixel boxcar and plotted in the rest frame,
   using a semi-automatic procedure, adapted from the IRAF task {\it splot}, to detect
   the \nev$\,\lambda$3426 feature in emission and compile a list of candidate AGN.
   This supervised but automatic preliminary analysis, which identified a possible
   \nev \ emission line only if its peak is above 2.5 times the significance level as
   derived by the noise of the surrounding continuum and its center is within
   $\pm20\,$\AA \ of the expected position, reduced to a few hundreds the number
   of galaxies to be further examined. The final phase of the selection process consisted
   of a careful examination of both the one-dimensional and two-dimensional
   sky-subtracted spectra, in order to: eliminate the spurious detections (sky line
   residuals, cosmic rays residuals, zero order contaminations); mark as secure
   candidates those galaxies with both the \nev \ emission lines clearly visible
   at the expected positions; and accurately verify the reliability of the 
   \nev$\,\lambda$3426 lines, when the three times fainter \nev$\,\lambda$3346
   feature is not detectable.      
   Following the laborious but accurate selection process presented above, we
   identified 95 zCOSMOS galaxies with a secure detection of \nev \ in their spectra.
   The permitted narrow emission lines visible in the observed spectral range of
   these galaxies (i.e., \mgii, \Hgamma, and \Hbeta) have been analyzed in order
   to investigate the possible presence of a faint broad component. We did not
   find a broad component in any spectrum of the \nev-selected objects,
   and we estimated an upper limit to the flux ratio of the narrow and broad
   components of the order of 10-20\%, depending on the emission line
   intensity and continuum S/N. Given these spectral properties, we
   can confidently classify all the selected objects as Type~1.9-2 AGN;
   hence the 95 galaxies constitute our bona-fide obscured AGN sample.
   
   In the redshift range where the \nev \ falls within our spectral coverage,
   the zCOSMOS survey discovered 112 broad line AGN. We applied the
   same analysis to this \tyi~AGN sample, detecting the \nev \ emission
   line in 45 objects (40\%).
%
   \begin{figure}
   \resizebox{\hsize}{!}{\includegraphics{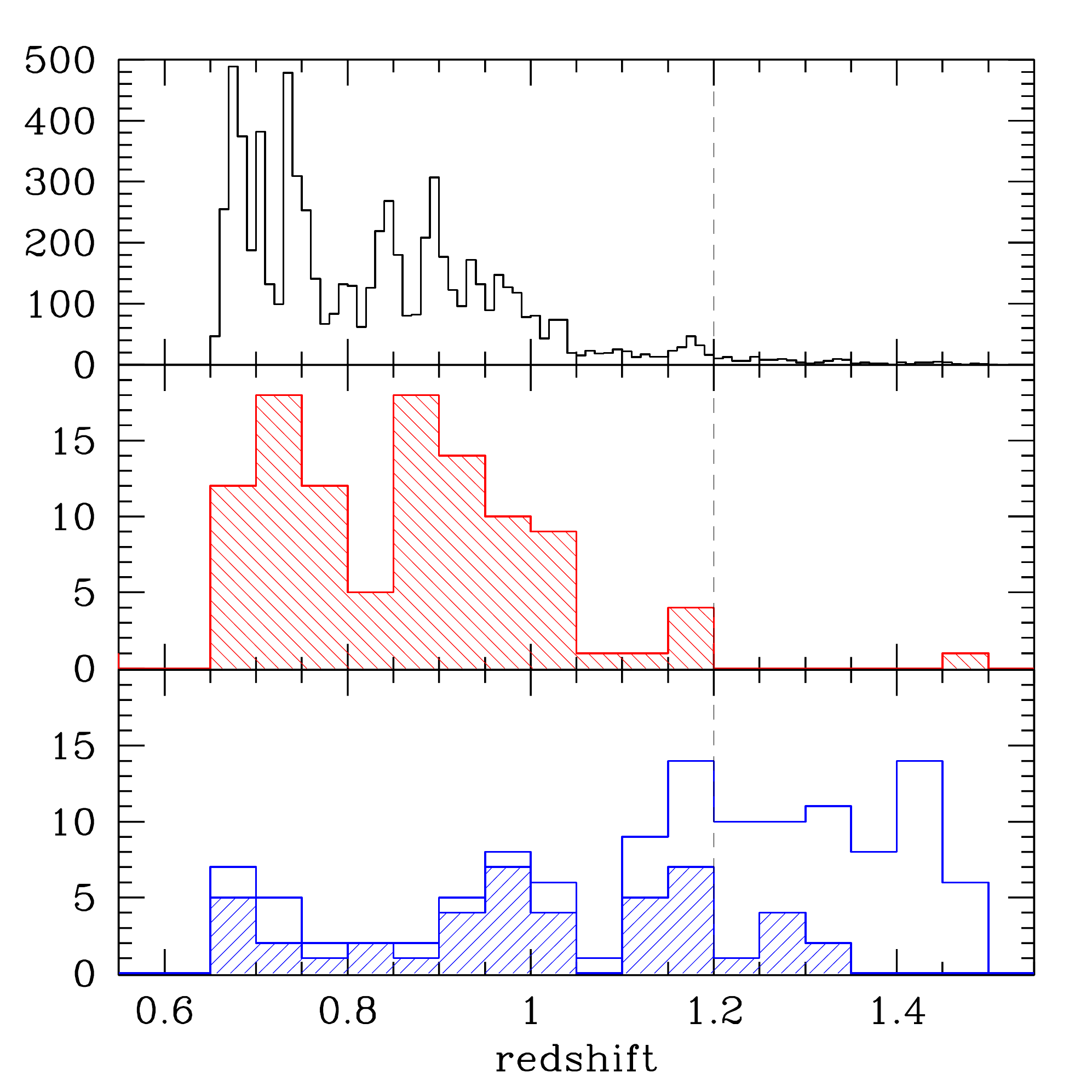}}
      \caption{Redshift distributions of zCOSMOS extragalactic objects in
      the redshift range examined (0.65$\,<\,${\it z}$\,<\,$1.50). Top panel: parent
      galaxy sample. Middle panel: narrow emission line galaxies with \nev \ detection.
      Lower panel: broad line AGN selected by zCOSMOS survey (empty histogram); 
      the hatched histogram shows the \nev \ detected objects. The vertical dashed line
      marks the upper redshift limit at z=1.2.
         \label{Figzdistrib}
      }
   \end{figure}

   The redshift distributions of our AGN samples, along with that of the parent galaxy
   population, are presented in Figure~\ref{Figzdistrib}. The redshift distribution
   of the \nev-selected emission line galaxies (the \tyii~AGN sample;
   middle panel) clearly follows that of the zCOSMOS galaxies, with a steady
   decline at redshifts greater than $\sim$0.9 and then an almost complete lack of
   objects above z=1.2, with a single noteworthy exception at z=1.45.
   The reason for the paucity of \nev-detected objects in the highest redshift interval
   is two-fold: first, the magnitude-limited zCOSMOS survey naturally culled normal
   galaxies, as well as obscured host-dominated \tyii~AGN, up to z$\sim$1.2
   \citep{OLF05, Lilly07}.
   Second, at redshifts greater then 1.2, the \nev \ feature enters the wavelength
   region where the zCOSMOS spectra are severely affected by fringing, making it more
   difficult to detect the emission line over the noisy continuum. As a confirmation of this
   effect, we can look at the \tyi~AGN sample, which was not selected on the basis
   of \nev-detection: at $z \le 1.2$, the fraction of broad~line objects with a detected
   \nev \ emission line is 62\% (37/60), whereas the ratio drops to a mere 13\% (7/52)
   among objects~with $z >1.2$. Due to the difficulties in identifying \nev-selected AGN
   above this redshift threshold, we limited the redshift interval to 0.65$\,<\,${\it z}$\,<\,$1.20,
   drawing two final samples of  94 narrow line (\tyii) AGN and 60 broad line (\tyi) AGN.
      
 \subsection{\tyii\ vs. \tyi\ Samples}
%
   \begin{figure*}
   \resizebox{\hsize}{!}{\includegraphics{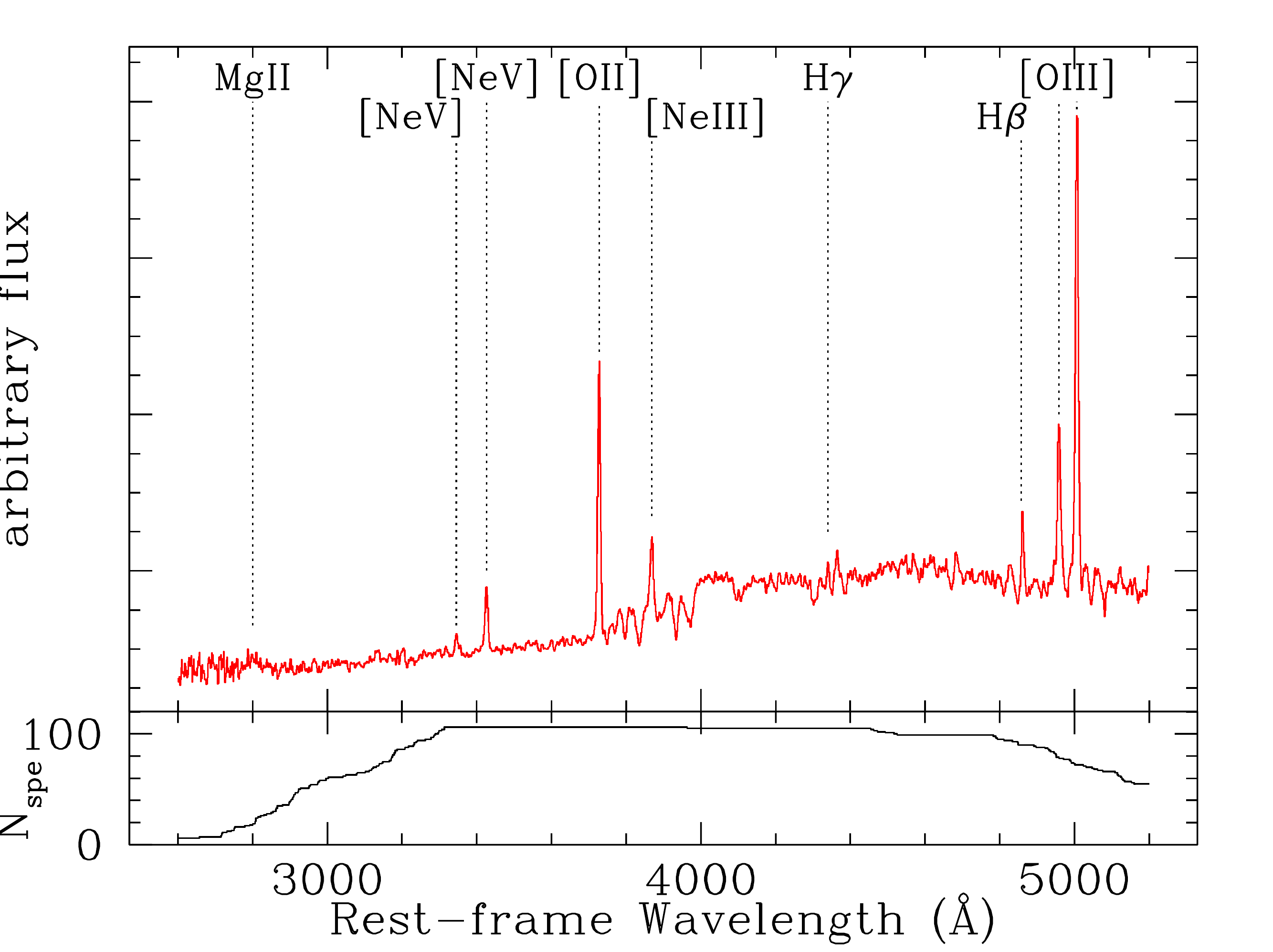}}
      \caption{Composite spectrum of the \tyii~AGN sample with identification of the main
      emission lines. The flux is per unit wavelength (F$_\lambda$), and normalization is 
      arbitrary. The number of single spectra contributing to the composite
      is shown in the bottom panel as a function of the rest-frame wavelength.  \label{compoNL}
      }
   \end{figure*}

 In this section we compare the mean optical properties of the two
 AGN samples, the \nev-selected \tyii\ sample and the \tyi\ sample, the latter 
 selected on the sole basis of the presence of  a broad component in the emission lines
 visible in the spectra (hereafter NL- and BL-samples). First, we generated
 composite spectra for each AGN class
 by averaging all the available zCOSMOS spectra included in that class. To create the
 composite, each spectrum was shifted to the rest-frame according to its redshift
 (with a 1.5\AA\ rest-frame bin\footnote{At the median redshift of the analyzed galaxy
 sample (z$\sim$0.82), a rest-frame bin of 1.5\AA\ matches the pixel size of the
 observations.}) and normalized to a common wavelength range, always
 present in the observed spectral window. An identical weight was assigned to each
 individual spectrum, to avoid biasing the final composite towards the brightest objects. 
 
 In Figure~\ref{compoNL} the average spectrum, obtained stacking the spectra of 
 ninety-four \nev-detected galaxies, is plotted. It clearly shows strong, high-ionization
 narrow lines, in particular those of Neon, but very faint \mgii\ emission line.
 In Figure~\ref{compoNL} the main emission lines visible in the rest-frame spectral
 range covered by the observations are also labelled, while a summary of the principal
 emission line parameters is presented in Table~\ref{smac}. Indeed, 
 the \tyii\ sample composite spectrum closely resembles the spectrum of a Seyfert~2 galaxy,
 although the line ratios would place it in the transition region of the diagnostic diagram
 involving the ``blue'' emission lines \citep[\oiii$\lambda$5007, \Hbeta, and 
 \oii$\lambda$3727;][]{Lamare04}, commonly used at these redshifts.
 All the emission lines are unresolved (the upper limit of 700~km/s roughly corresponds
 to the spectral resolution of the zCOSMOS data), with the possible exception of the \nev\ line
 itself, which seems marginally resolved, and of the very faint \mgii\ feature. But in the
 latter case the low S/N of the line prevents us from obtaining a reliable FWHM estimate.

 In order to obtain accurate flux measurements of the narrow Balmer emission
 lines in the spectral range, we used \citet{BC03} (BC03) population synthesis models 
 to fit and subtract the stellar continuum in the composite spectrum of the \tyii~AGN;
 the fitting procedure allowed us to account for the underlying stellar absorption of
 the Balmer lines. The mean optical extinction in the Narrow Line Region (NLR),
 derived from the observed \Hbeta/\Hgamma\ flux ratio, is quite low
 (\hbox{$\langle$E(B$-$V)$\rangle\,$=$\,0.18$}). This is not unexpected, since the
 AGN obscuration occurs in the inner core of the galaxies, and a significant amount of
 dust in the NLR would have prevented us from detecting the blue \nev\ emission line. 
 The absorption-line continuum of the composite could also provide useful information
 about the average stellar content of the galaxies hosting our \tyii~AGN sample.
 A set of 39 BC03 template spectra, spanning a wide range in age and metallicity,
 have been used to fit the emission-line-free regions of the composite. The best-fit model
 corresponds to an old stellar population with an exponentially declining star-formation,
 with the ratio between galaxy's age and SF e-folding time of t/$\tau$$\approx$2.
 These model parameters roughly approximate a galaxy of Hubble Type Sa/Sb \citep{Buzz05}.
 
\begin{figure}
   \resizebox{\hsize}{!}{\includegraphics{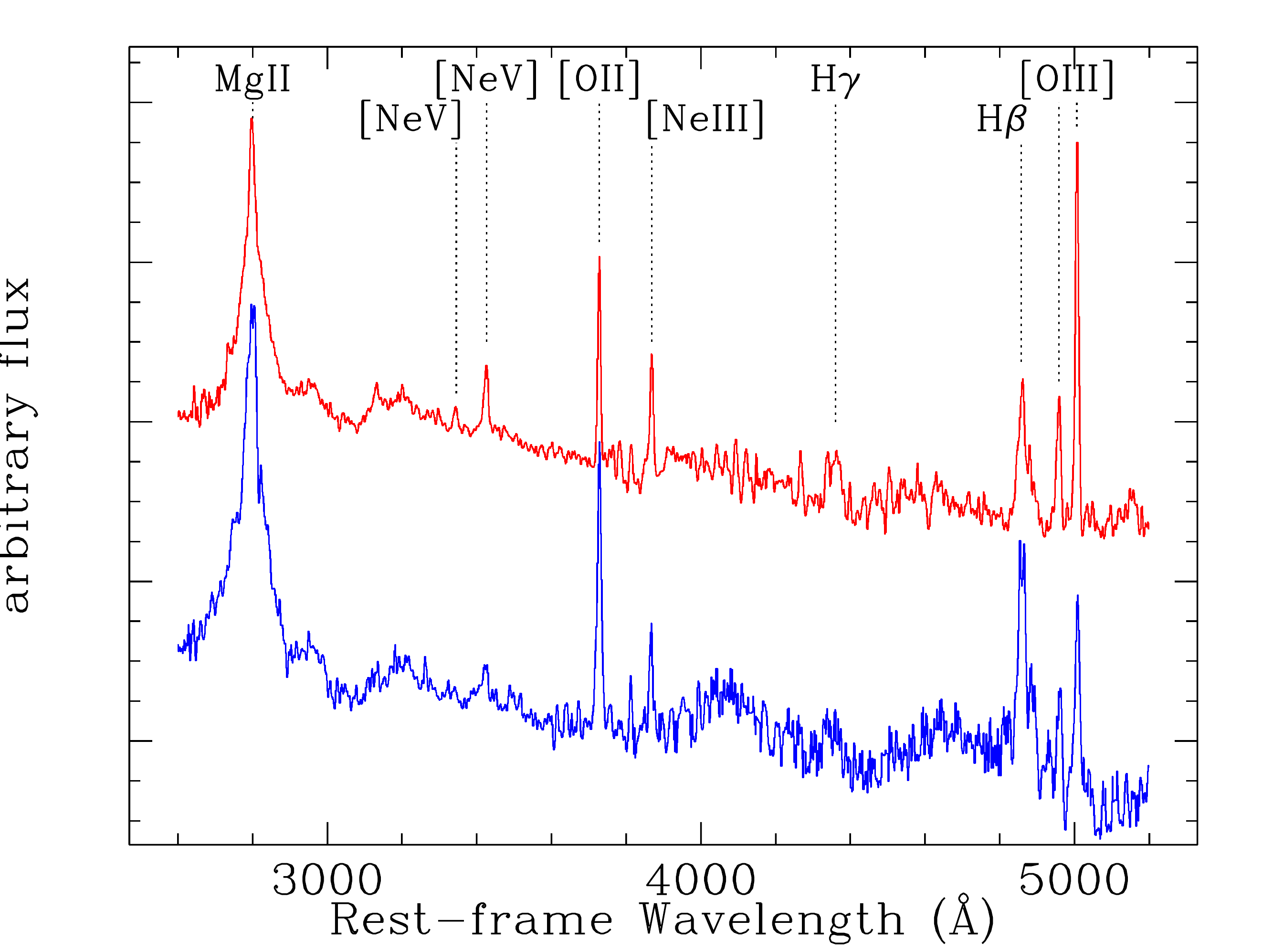}}
      \caption{Average spectra of the \tyi~AGN samples with the identification
      of the main emission lines: the lower composite spectrum (in blue) represents
      the \tyi~sample without \nev \ detection,
      while the upper composite spectrum (in red) is the average of the \nev-detected \tyi~AGN;
      the spectra are offset for clarity. The flux is per unit wavelength (F$_\lambda$), and normalization
      is arbitrary. \label{compoBL}
      }
   \end{figure}

 We also computed the average spectra of \tyi \ AGN, by stacking spectra
 the thirty-seven broad line AGN with a \nev~detection and the twenty-three
 objects without a detectable \nev \ line. 
 The two composite spectra are presented in Figure~\ref{compoBL}, showing evident
 similarity. Nevertheless, looking at Table~\ref{smac}, where the principal emission line
 parameters are listed, it is possible to identify a main difference: 
 the \Hbeta/\Hgamma \ flux ratio in \tyi~AGN without \nev \ emission is
 significantly larger than in \nev-detected ones, suggesting a larger mean optical
 extinction in the corresponding sub-sample. 
 The higher average extinction in the sample 23 BL-AGN without \nev-detection
 could explain why this feature is missing in the individual spectra,
 but visible in the high S/N stacked spectrum.
 
%
   \begin{table}
      \caption[]{Spectral measurements in AGN composite spectra.}
         \label{smac}
     $
     \resizebox{0.49\textwidth}{!} {         
       \begin{tabular}{lrcrrrr}
            \hline
            \noalign{\smallskip}
            &\multicolumn{2}{c}{NL-AGN}   &\multicolumn{2}{c}{BL-AGN}  
            &\multicolumn{2}{c}{BL-AGN} \\
           &&&\multicolumn{2}{c}{(\nev-det.)}&\multicolumn{2}{c}{(no-\nev)} \\
            \noalign{\smallskip}
            line & EW & FWHM  & EW & FWHM  & EW & FWHM \\
             & [\AA] & [km/s] & [\AA] & [km/s] & [\AA] & [km/s] \\
            \hline
            \noalign{\smallskip}
            ~\mgii{$\lambda$2800} & 6.2 & & 42.0 &$\sim$5000 & 64.0 &$\sim$7500 \\
            \nev{$\lambda$3346} & 2.3 & & 0.9 & & 0.6 & \\
            \nev{$\lambda$3426} & 8.2 & $\sim$900 & 3.1 & $\sim$1000 & 1.4 & $\sim$1100 \\
             \oii{$\lambda$3727} & 29.3 & $<$700 & 7.0 & $<700$ & 12.8 & $\sim$900 \\
             \neiii{$\lambda$3869} & 6.0 & $<$700 & 3.5 &  $<700$ & 3.1 & $<$700 \\
             ~\Hgamma & 3.2 & $<$700 & 8.6 & $\sim$2500 & 3.8 & $\sim$2000 \\
             ~\Hbeta & 8.2 & $<$700 & 21.4 & $\sim$2000 & 22.0 & $\sim$2000 \\
             \oiii{$\lambda$4959} & 9.0 & $<$700 & 6.9 & $<$700 & 4.2 & $\sim$800 \\
             \oiii{$\lambda$5007} & 33.2 & $<$700 & 20.5 & $<$700 & 13.1 & $\sim$800 \\
            \noalign{\smallskip}
            \hline
         \end{tabular}
     }
     $
  \tablefoot{All values are rest-frame. \Hgamma \ is severely contaminated by 
the adjacent [O\,\textsc{iii}]\,$\lambda 4353$ line in BL-AGN composite
spectra. The FWHM of low S/N emission lines is highly uncertain, and its value
is missing in the table; FWHM upper limits refer to unresolved emission lines.}
   \end{table}

\begin{figure}
   \resizebox{\hsize}{!}{\includegraphics{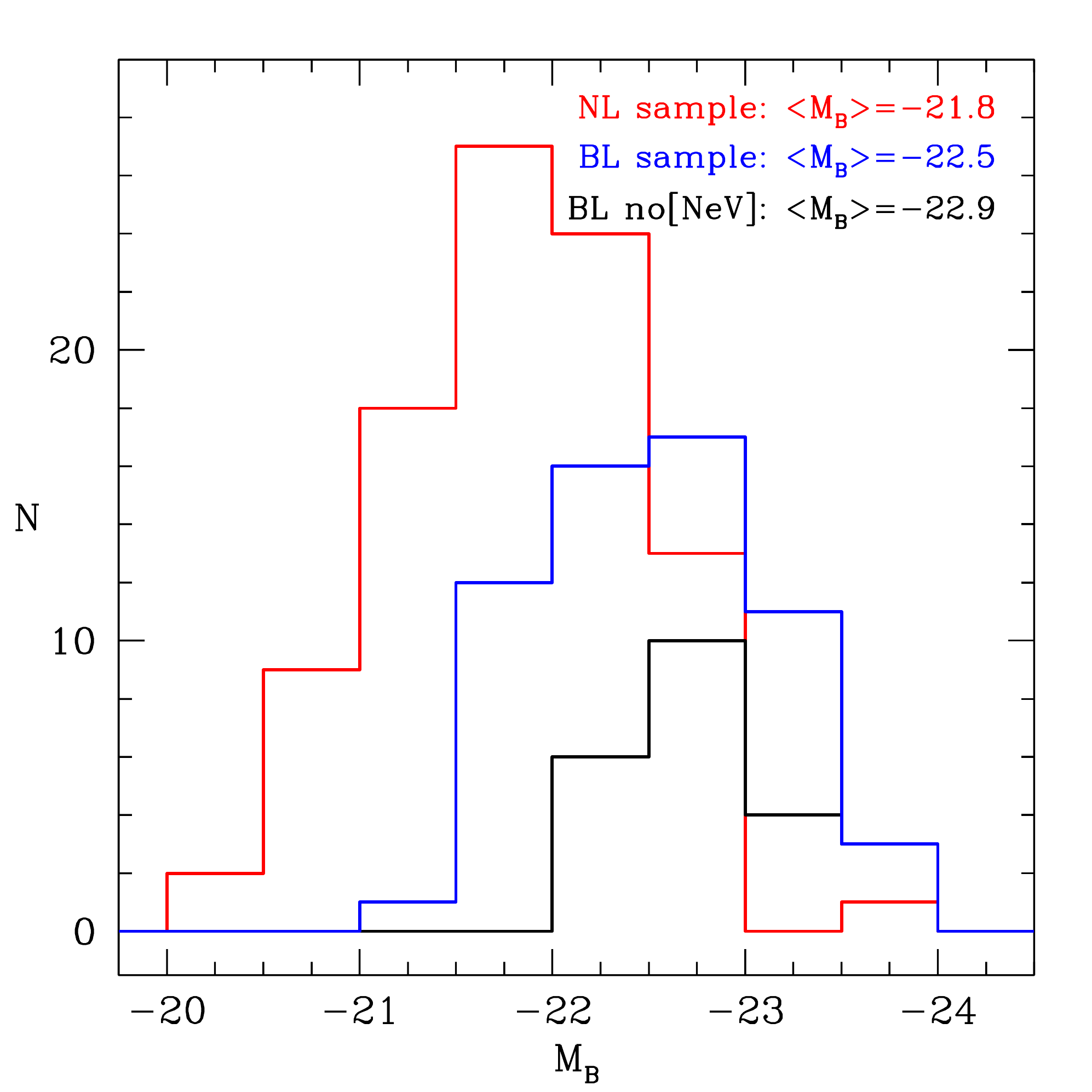}}
      \caption{Absolute B magnitude distributions: the red histogram represents the sample
      of  94 \tyii~AGN, the blue histogram the sample of 60 \tyi~AGN, and the
      black histogram the sub-sample of 23 BL-AGN with undetected \nev.
      The median absolute B-magnitudes of the three samples are also indicated.
         \label{FigMbdistrib}
      }
   \end{figure}

 The non-detection of the \nev\ emission line in a fraction of \tyi~AGN spectra
 is also probably related to the presence of a stronger continuum
 from the central engine in these objects, continuum that is shielded in the NL-sample.
 In support to this hypothesis, the absolute B-magnitude distributions of the AGN samples
 are presented in Figure~\ref{FigMbdistrib}. Absolute magnitudes were computed following
 the method described by \citet{Ilb05}, using a set of templates for the K-correction and 
 all the available photometric information. In order to reduce the importance of the
 K-correction assumption, rest-frame absolute magnitude were derived using
 the apparent magnitude from the closest (depending on the redshift) observed band.
 In Figure~\ref{FigMbdistrib} a clear trend is visible: the median absolute B-magnitudes
 in the three sample distributions are -21.8, -22.5 and -22.9, respectively, for the total
 NL-sample, the total BL-sample, and the sub-sample of unobscured AGN with undetected
 \nev. It is indeed expected that the NL-sample is composed by less
 luminous objects with respect to the unobscured AGN which belong to the BL-sample, but
 the higher luminosity of the BL-AGN with undetected \nev \  
 corroborates the idea that the detection of this faint spectral feature becomes more
 difficult when the (unobscured) nuclear continuum increases. Further analysis of the
 \tyii \ sample completeness, in terms of \nev \ line detection, is discussed in the next
 section.

%
%

%

 Finally, we compare the properties of the \hbox{average spectrum} of our BL-sample with 
 those of the SDSS quasar composite \citep{VB_SDSS01}: the \mgii \ broad emission
 line is significantly stronger, on average, in our AGN sample than observed in the more
 luminous SDSS quasars \citep[EW=32\AA; see Table~2 in][]{VB_SDSS01}. This is expected in
 the framework of the ``Baldwin Effect'' \citep[][but see also a thorough discussion of the
 observational biases in Zamorani et~al. 1992]{Bald77}. Conversely, the \Hbeta \ relative intensity
 is lower in our composite spectra than in the SDSS one (EW$\approx$22\AA\ and 46\AA,
 respectively), and the continuum is significantly redder. The latter results are consistent
 with the hypothesis that, in the red part of the rest-frame optical range, our composite spectra 
 are significantly contaminated (and reddened) by the host galaxy stellar light. Very similar results
 were obtained by \citet{Gav06}, who selected their \tyi~AGN sample from the VVDS survey,
 a spectroscopic survey which shares many characteristics (magnitude limit, instrument)
 with zCOSMOS. Although a detailed study of the broad line AGN component in the
 zCOSMOS survey is beyond the scope of this paper, and will be addressed in a future work,
 the preliminary analysis presented above indicates that the zCOSMOS BL-samples, both
 the \nev-detected sub-sample and the no-\nev\ one, do not show atypical spectral
 properties when compared to other known quasar samples.
 \subsection{\tyii~AGN Sample Completeness}
 
 The principal aim of this paper is to select and study a sample of obscured AGN at
 redshift $\approx$ 1. The BL-sample has been collected mainly for comparison purposes.
 For this reason, our main concern is related to the efficiency of our detection technique
 in selecting narrow line \nev-emitting galaxies, having already interpreted the non-complete
 detection of the \nev\ emission line in \tyi\ objects as a combined effect of smaller emission
 line EWs and noise in zCOSMOS spectra.
 
\begin{figure}
   \resizebox{\hsize}{!}{\includegraphics{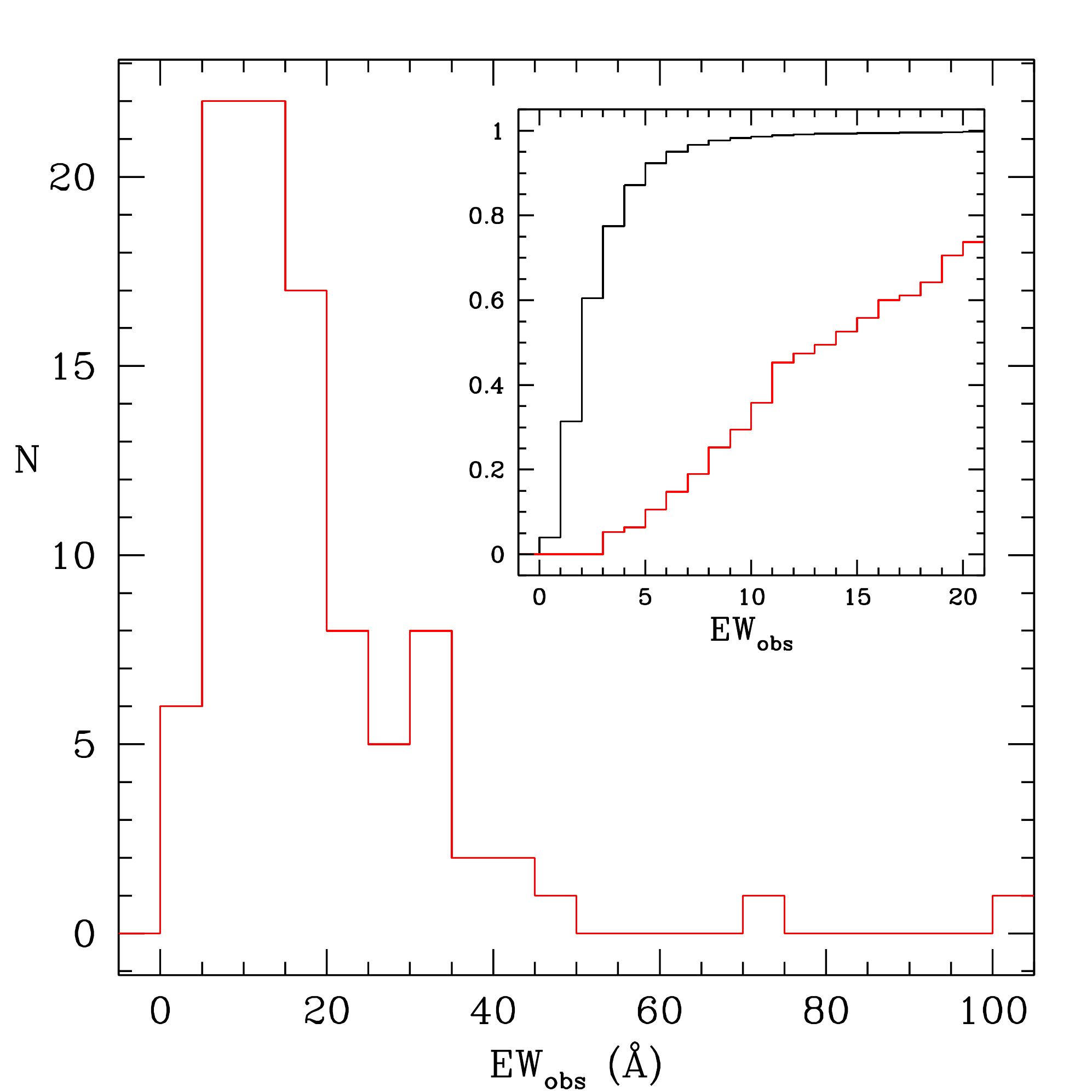}}
      \caption{Observed \nev~Equivalent Widths distribution of the sample of \tyii~AGN.
      The inset shows (in red) the correspondent cumulative distribution of
      the detected \nev\ emission lines, along with the cumulative distribution (in black)
      of the EW upper limits for the zCOSMOS galaxies with no \nev\ detection.
         \label{Figewdistrib}
      }
   \end{figure}

 In order to support the effectiveness of our technique in selecting a fairly complete sample of
 \tyii~AGN, at least in terms of \nev\ emission, we analyze the distribution of the detected
 line EWs, presented in Figure~\ref{Figewdistrib}. The observed EW distribution of the
 \nev\ emission line is peaked at around 10\AA, with a long tail up to $\approx$~100\AA.
 The inset of Figure~\ref{Figewdistrib}
 shows the cumulative EW distribution of the detected \nev\ lines, along with the
 cumulative distribution of the EW upper limits for the 7265 zCOSMOS galaxies with
 undetected \nev. The detection limits have been estimated from the S/N in the continuum
 adjacent to the line, following the procedure outlined by \citet{Migno09} on the basis
 of a larger sample of emission line galaxies drawn from the zCOSMOS survey.
 Comparing the two cumulative distributions, we find that 92\% 
 of the galaxies with undetected \nev\ have upper limits lower than 5\AA, 
 while only six \tyii~AGN show EW smaller than the same threshold. 
 Since there is no reason to expect that galaxies with loose upper limits
 include a different percentage of AGN than the parent sample (1.3\%),
 we can confidently conclude that our selection misses very few 
 strong \nev\ emitters (i.e. with EW larger than $\approx\,$5\AA).

\begin{figure}
   \resizebox{\hsize}{!}{\includegraphics{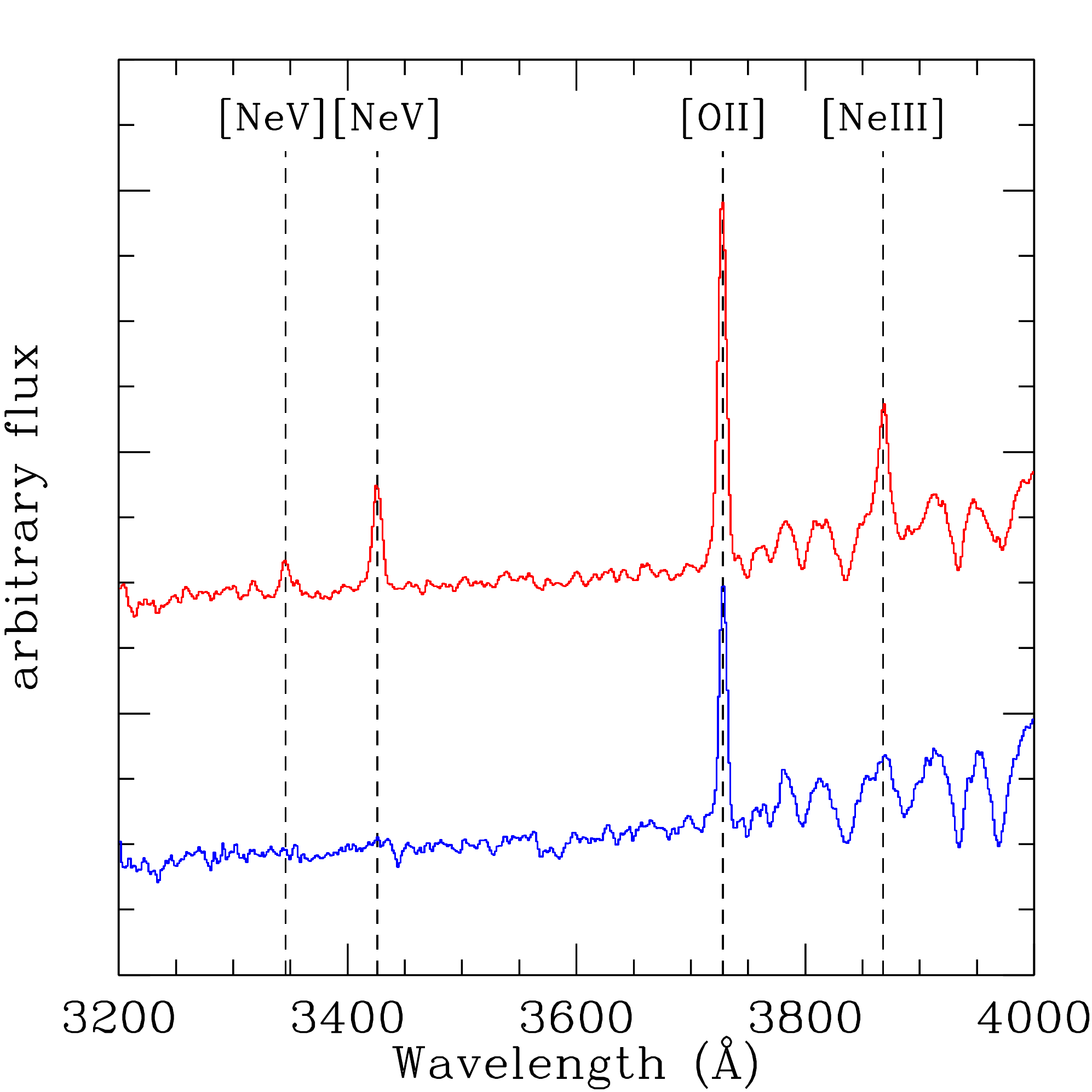}}
      \caption{Composite spectra of the \nev-selected, \tyii~AGN sample (top, in red) and of the
      X-ray emitting galaxies without \nev\ detection (bottom, in blue);
      the spectra are offset for clarity.
        \label{Figcompa}
      }
   \end{figure}

 We further examine our detection efficiency exploiting the multi-wavelength data set
 of the COSMOS field, using the Chandra-COSMOS Survey \citep[C-COSMOS;][]{Elv09}
 to pinpoint AGN candidate in X-rays. The sample of zCOSMOS galaxies without
 broad lines in their spectra and 0.65$\,<\,${\it z}$\,<\,$1.20, includes
 5148 objects inside the Chandra mosaic; 180 of these galaxies have been associated
 with an X-rays source in the C-COSMOS catalog 
 (details on the X-rays catalog and matching criteria can be found in 
\citealt{Puccetti09} and Section 5.1).
 Most of these galaxies likely harbor an active galactic nucleus, since their X-ray luminosity
 is larger than $10^{42}$~erg/s, and it has been shown that other sources of X-ray
 emission in normal galaxies \citep[i.e. high-mass binaries;][]{Moran99} 
 cannot easily account for such high luminosities. 
 We stacked 158 optical spectra of the X-ray emitting galaxies without \nev \ detection using
 the same recipe adopted for the \nev-detected \tyii\ AGN sample, and the two composite
 spectra are shown in Figure~\ref{Figcompa}. Visually comparing the two spectra, it is striking
 the complete absence of the \nev \ doublet in these X-ray emitting galaxies;
 an upper limit of 0.5\AA \ to the \nev\ equivalent width can be obtained from
 the S/N of the continuum. 
 Therefore, we can confidently assert that the selection process should not have
 missed a significant number of galaxies with the \nev \ emission line in the EW range
 of 1-3\AA, since such a population of weak \nev \ emitters do not emerge even
 in a complementary X-ray selected AGN sample. On the other hand,
 the presence of a significant number of X-ray emitting galaxies, probably hosting an
 obscured AGN, without a detectable \nev \ emission, clearly indicates that this
 technique alone is not sufficient to select a complete sample of \tyii~AGN. The mean
 optical extinction in these X-ray emitting galaxies, derived from the \Hbeta/\Hgamma\
 flux ratio measured in their composite spectrum, is \hbox{$\langle$E(B$-$V)$\rangle\,$}=$\,0.27$,
 larger than in the \nev-selected \tyii~AGN sample, suggesting dust extinction on galactic
 or NLR scale as a possible cause for the \nev \ paucity in these objects.

%
%

%
%


\section{The \nev-selected \tyii~AGN sample properties}

  \subsection{Stellar Mass distribution of the host galaxies}

 Stellar masses, for both the \tyii~AGN and the galaxy parent samples, have 
 been computed by \citet{Bolzonella10}. Briefly, they used different
 stellar population synthesis models to fit the large set of optical and
 near-infrared photometry available in the COSMOS field, using a $\chi^2$ minimization
 to find the best-fit model, at a fixed redshift $z = z_{spec}$ from the zCOSMOS
 survey. Further details on stellar masses' determination can be found in
 \citet{Bolzonella10} and \citet{Pozzetti10}.
%
%

 Previous studies of \tyii~AGN host galaxy properties have assumed that the
 contamination of the AGN light to their stellar masses measurements is
 negligible \citep{Silverman08, Schaw10}. 
 \citet{Bongiorno12} analyzed the differences between the stellar masses
 computed using SED fitting with only a galaxy component and 
 those obtained with two components (AGN and galaxy). They applied the
 two SED fitting to a large sample of AGN in the COSMOS field:
 our \nev-selected AGN, included in their work, shows the least differences
 in the computed stellar masses, between the various AGN classes, with
 a median of mass ratios \hbox{$M_{(gal)}/M_{(gal+AGN)}=1.04$}.
 Our \nev-selected \tyii~AGN sample is composed by galaxies
 hosting relatively low-luminosity and obscured AGN, so we can
 confidently compare their derived properties with those of the parent sample. 
%
   \begin{figure}
   \resizebox{\hsize}{!}{\includegraphics{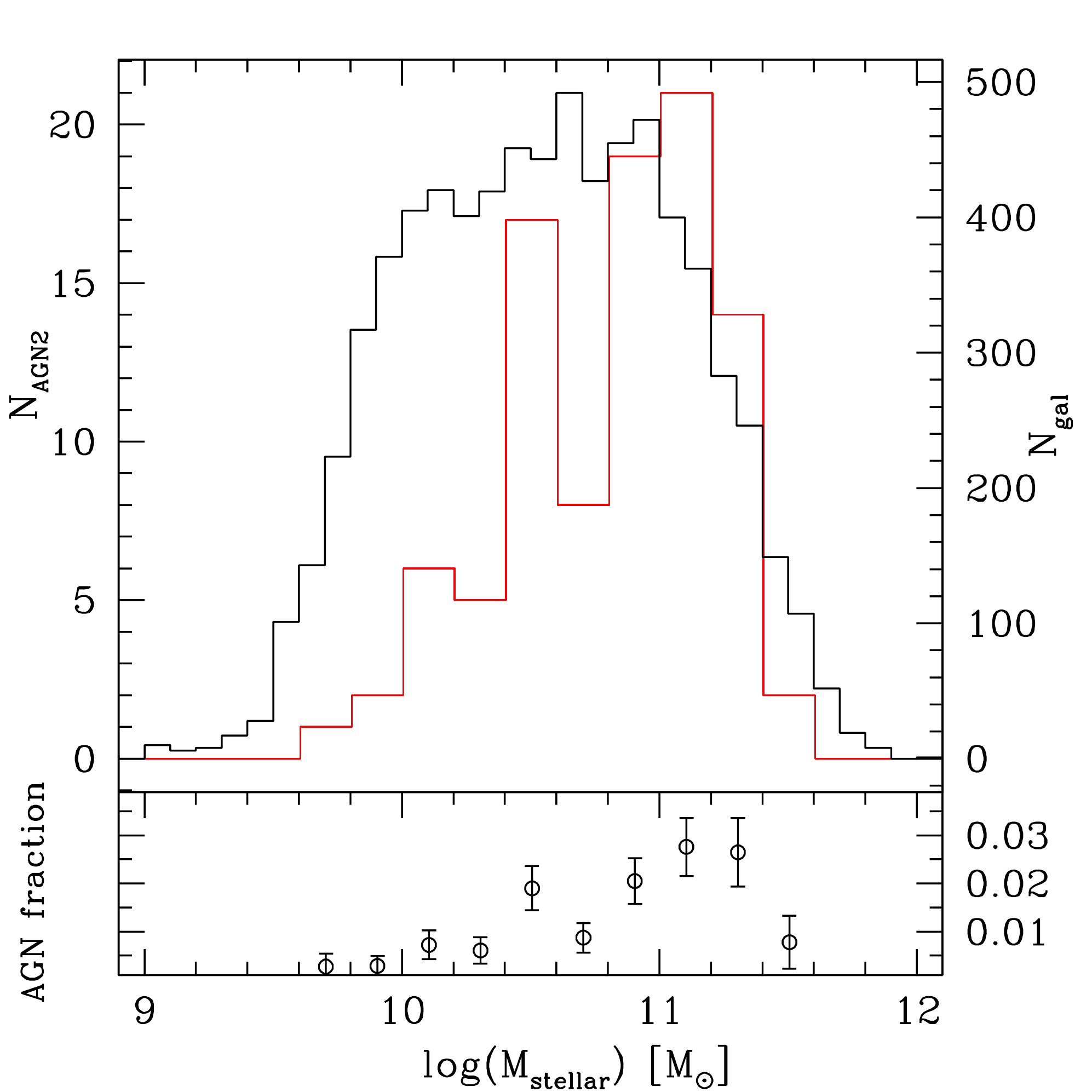}}
      \caption{Host stellar mass distribution for the \tyii~AGN sample (red histogram)
      compared with that of the galaxy parent sample (black histogram; 
      0.65$\,<\,${\it z}$\,<\,$1.20). The y-axes used for the two histograms are scaled relative
      to each other in order to ease the comparison.
      In the bottom panel, the ratio between the two populations is shown. 
         \label{FigMdistrib}
      }
   \end{figure}

 Figure~\ref{FigMdistrib} shows the stellar mass distribution of galaxies hosting the
 \tyii~AGN sample (red histogram), along with that of the parent galaxy sample (black histogram). 
 From the top panel of Figure~\ref{FigMdistrib}, we can see that the stellar mass
 for most \tyii~AGN hosts is in the range $1\times 10^{10}-\;3\times 10^{11} M_\odot$,
 while the galaxies of the parent sample have masses down to $5\times 10^9M_\odot$.
 The median stellar mass of the \tyii~AGN host galaxies is $8\times 10^{10}M_\odot$, while
 for the parent galaxy sample is $3.7\times 10^{10}M_\odot$. 
 We used a two-populations Kolmogorov-Smirnov (K-S) test to assess the significance
 of the difference between the masses of the parent sample and the AGN host galaxies,
 finding that they differ at very high significance ($> 5\sigma$).

 The fraction of \nev-selected obscured AGN in our zCOSMOS sample is shown
 as a function of the host stellar mass in the lower panel of Figure~\ref{FigMdistrib},
 with the percentage of \tyii~AGN rising from less than 1\% at $\sim 2\times 10^{10}M_\odot$
 to around 3\% at $\sim 2\times 10^{11}M_\odot$.
 A similar trend is shown by purely X-ray selected AGN in zCOSMOS
 \citep{Silverman09}, and in larger X-ray sample 
 \citep{Aird12}, although we will show below that the two selection techniques
 (\nev \ and X-ray) are more complementary than overlapping. Also in optically
 selected AGN samples the detection rate is strongly dependent
 on host galaxy mass, as established in a sample of SDSS emission-line
 galaxies \citep{Kauff03}.

  \subsection{Morphologies of the host galaxies}

 In order to compare the morphologies of our AGN host galaxies to those of normal galaxies,
 we built a control sample of galaxies from the zCOSMOS survey. Given the different
 mass distribution highlighted in Figure~\ref{FigMdistrib}, for each of the 94 narrow line AGN
 we selected 8 galaxies from the parent sample with matched redshift and stellar mass.
 We made use of the morphological catalog by Nair et~al. (in preparation)
 that, using the F814W-band images from the Advanced Camera for Surveys (ACS)
 available in COSMOS and following the method applied in \citet{Nair10},
 visually classified all the galaxies belonging to zCOSMOS, and divided
 them into different types (Ell, S0, Sa-Sd, and Irr).
 A clear classification in one of the Hubble types was possible for
 71 out of 94 (76\%) of the AGN host galaxies, while a similar fraction
 of galaxies (597/752, 79\%) have been classified in the control sample.
 The objects which do not have a regular morphology assigned to them are
 mainly of two classes: for some galaxies, the resolution and signal-to-noise
 is such that it was not possible to establish an accurate morphological type,
 but the majority of  the unclassified objects shows peculiar morphologies
 (i.e. doubles, tadpoles and chain galaxies) and do not
 find a natural place in the Hubble sequence.

\begin{figure}
   \resizebox{\hsize}{!}{\includegraphics{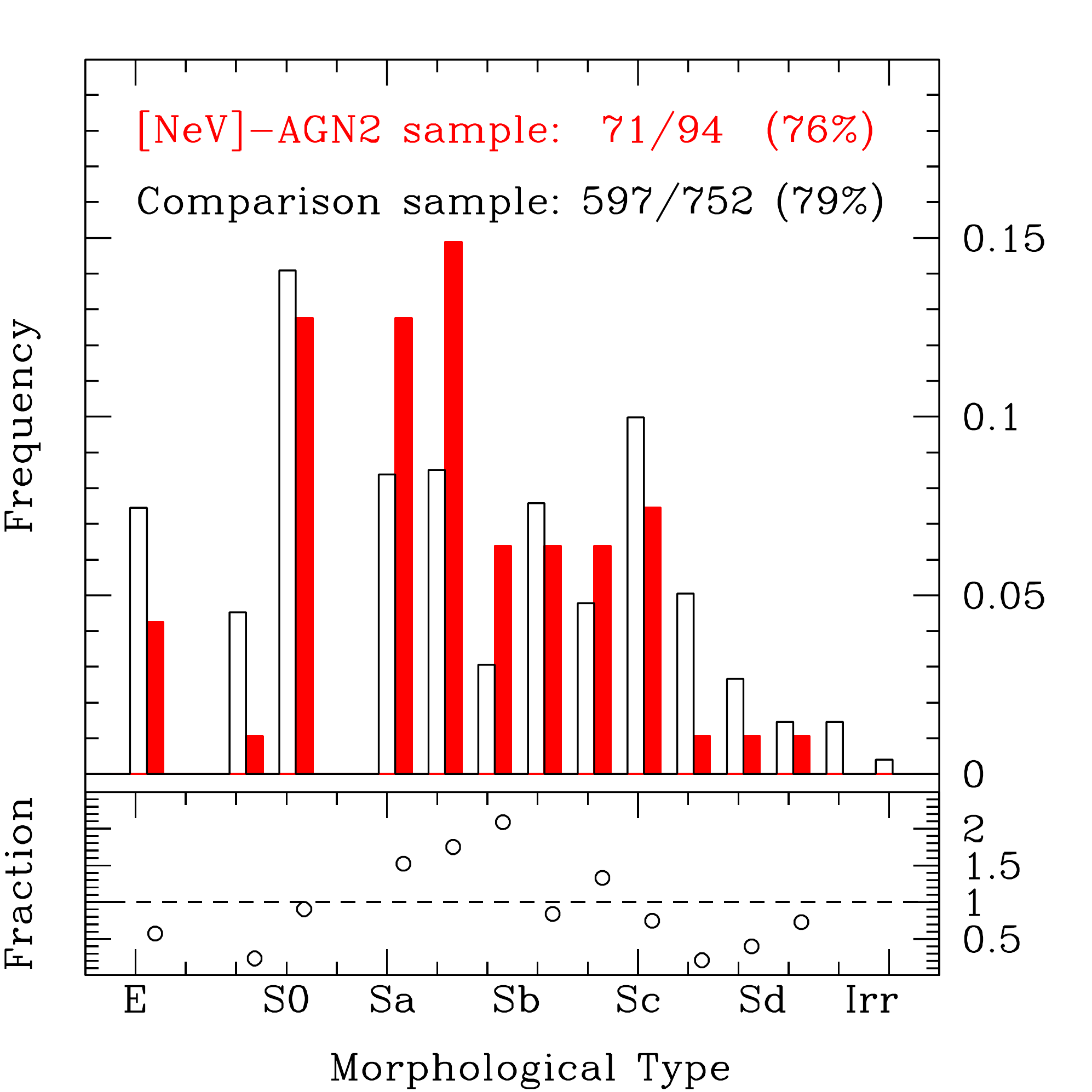}}
      \caption{Morphology distribution of the 71 classified \tyii \ AGN hosts (red
      filled histogram) compared with that of the 597 normal galaxies belonging to the control
     ($8\times$)~mass-matched sample (black empty histogram). In the bottom panel, the
     fraction between the relative frequencies of the two populations is plotted as a
     function of the Hubble types. 
         \label{FigMorph}
      }
   \end{figure}

 The AGN host galaxy morphology distribution is plotted along with
 that of the control sample in Figure~\ref{FigMorph}, where only objects with
 assigned Hubble type are shown. The two distributions are not strikingly
 different, since the significance of the difference between the two populations
 is at about 2~$\sigma$ level, as estimated from a K-S test.
 Even if the two distributions are similar, the lower panel of  Figure~\ref{FigMorph}
 seems to suggest a difference in the trend of the relative frequency of the
 Hubble types:  the hosts of the \nev-selected \tyii~AGN prefer the early-spirals
 morphologies (Sa-Sb), with an expected paucity of late-spirals and Irregulars, but
 also with a lower fraction, with respect to the control sample, of elliptical galaxy hosts.
 
%
%

 \subsection{Optical spectroscopic properties}

%
   \begin{figure}
   \resizebox{\hsize}{!}{\includegraphics{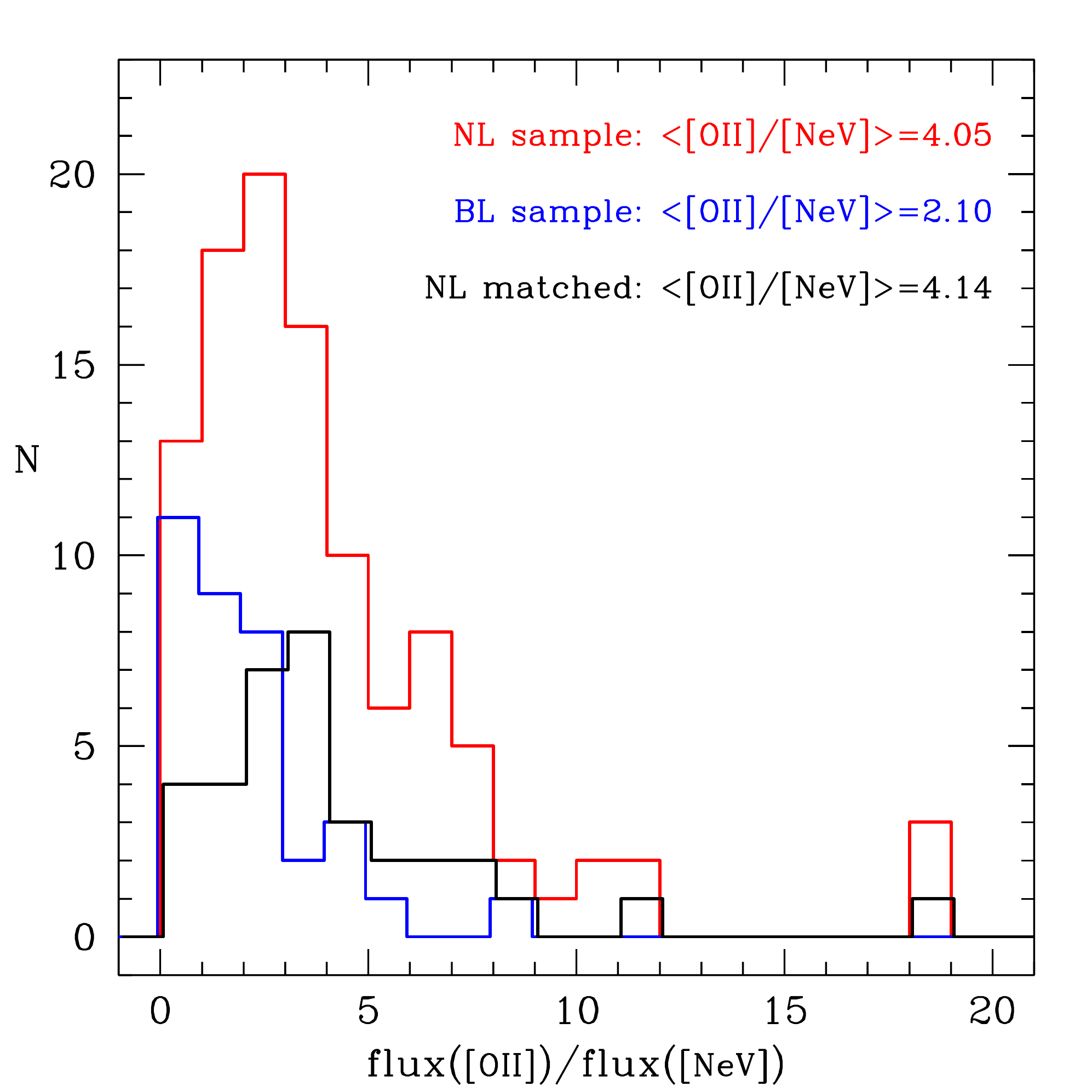}}
      \caption{\oii/\nev \ line flux ratio distributions: the red histogram represents the sample
      of  \tyii~AGN, the blue histogram the sample of \tyi~AGN with detected \nev, and
     the black histogram the sub-sample of NL-AGN matching the redshift/flux characteristics
     of the BL-AGN. The average line flux ratios of the three samples are also indicated.
         \label{FigO2Ne5}
      }
   \end{figure}

We measured the \nev \ and \oii \ emission lines in all the available zCOSMOS spectra
of the AGN included in both the \tyi \ and \tyii \ samples selected by \nev-detection.
We used a semi-automatic procedure that exploits the IRAF task $splot$: first, the continuum 
was automatically fitted to fixed wavelength intervals, although our procedures also
enabled interactive adjustments of the continuum level to improve the line measurement
in noisy spectra. Then, equivalent widths (EWs) and fluxes were measured by applying a
Gaussian-function fitting algorithm of the continuum-subtracted line profiles.
The measurement errors, estimated from the rms of the continuum
close to the line after the Gaussian fit is subtracted,
are of the order of 5-10\% for the \oii, and 10-20\% for the fainter \nev. 
Moreover, repeated observations exist for more than 5\% of the zCOSMOS sample,
and spectral measurements obtained from these
repeated spectra are consistent within the errors quoted above.

%

We investigate the \oii/\nev \ flux ratio in our \nev-selected
samples of broad and narrow line AGN. According to the unification model
\citep{Anto93}, the emission properties of the Narrow Line Region
should be similar in both AGN types. Thus, under the hypothesis
that the \nev \ emission and NLR \oii \ component
are linked, a larger line flux ratio in the obscured population,
with respect to typical \tyi~AGN values, would favor the scenario of enhanced
star formation in \tyii~AGN invoked by semi-analytic models of galaxy
formation and evolution \citep[][and references therein]{Gilli10}.

The mean (median) flux ratio is 4.05 (3.13) in the NL sample and
2.10 (1.62) in the BL sample. Because of different selection efficiencies
in the two samples (see section 3.2), we also extracted, from the larger
sample of NL-AGN,  a sub-sample with redshift and \nev \ flux
distributions matching those of the BL-AGN. Again, the mean (median)
\oii/\nev \ intensity ratio is 4.14 (3.51) in this BL-matched \tyii~AGN sample.
We measured the flux ratio also in the composite spectra of
\tyii~AGN and \nev-detected \tyi~AGN samples presented in section~3.2,
obtaining similar values of 3.62 and 1.94, respectively. 
The composite spectrum of the \tyii~AGN sample, selected at
redshifts 0.3$<$z$<$0.83 from the Sloan Digital Sky Survey, shows an
intensity ratio in good agreement with the averages measured in this work
\citep[4.5$\pm$0.3;][]{Zakamska06}.
The significantly larger value of the \oii/\nev \ flux ratio is therefore
suggestive of enhanced star formation, for a given \nev \ luminosity,
in the obscured population, at least with respect to the sample of \tyi~AGN. 

\section{Comparison with other \tyii~AGN selection techniques at {\textsl z}$\sim$1}

 While the cosmological evolution of unobscured QSOs has been relatively
 well studied out to very high redshifts ($z\sim5-6$) thanks to
 large optical surveys, mostly the 2dF QSO Redshift Survey
 \citep{Croom04} and the Sloan Digital Sky Survey \citep[SDSS;][]{SDSS06},
 there are not many samples of optically selected, high-z \tyii~AGN
 available in the literature, and their space density beyond the local
 universe is poorly known.  
 The selection of complete samples of obscured AGN is a difficult task,
 and the two methods commonly regarded as the most complete, the optical
 emission line selection and the X-ray detection, have their own biases.
 In the optical, dust extinction throughout the host galaxy
 can significantly reduce the observed emission line luminosity and/or alter
 the line ratios. Moreover, the emission line selection suffers from a
 redshift-dependent completeness, since at different redshifts the diagnostic
 diagrams exploit different line ratios, with different efficiencies and
 specificities. On the other hand, the X-ray selection, while largely unaffected
 by extinction in the host galaxy, is biased against sources in which nuclear
 emission is absorbed by Compton-thick gas close to the central engine.

 Consequently, the \tyii~AGN detection rate strongly depends on
 the adopted selection criteria and sample definition.
 There is no single known method that can select a complete
 sample of obscured AGN, and we are not aware of any selection
 technique capable to identify all the objects found by other methods. 
 In this section, we compare the result of our \nev-selection technique
 with other AGN selection methods, using as reference the parent sample
 of 7358 galaxies with 0.65$\,<\,${\it z}$\,<\,$1.20 and $I<22.5$, 
 selected from the zCOSMOS Redshift Survey.

  \subsection{X-ray selection}

The sample of zCOSMOS galaxies without broad lines in their spectra and redshift
range 0.65$\,<\,${\it z}$\,<\,$1.20 includes 5148 objects within the
Chandra mosaic; 180 of these galaxies show X-ray emission (all above
10$^{42}$~erg~s$^{-1}$) and are likely obscured AGN. 
%
%
Only 23 \nev-selected sources are detected by Chandra
within 1.2~arcsec from the optical position; the median distance between
the X-ray and optical counterpart is 0.46 arcsec.
To derive rest-frame, observed (i.e., prior to absorption correction)
2-10 keV fluxes, we have extracted the X-ray counts for these 
sources using the multiple pointings of the Chandra-COSMOS mosaic. 
Here we present a summary of the adopted procedure; 
a more exhaustive description is presented in a forthcoming paper
(Vignali et al., in preparation).

Because of the tiling strategy in the C-COSMOS field, every source can be
observed in more than one pointing and at different locations within
the \hbox{ACIS-I} field-of-view, hence the source count distribution is
generally characterized by different PSF sizes and shapes.
To properly account for all of these effects, we have used the
{\sc Acis Extract} software \citep{Broos10}, which extracts the source
counts from each observation using the 90\% of the encircled energy
fraction at 1.5 keV at the source location and then correcting for
aperture. 
X-ray photometry in the observed band corresponding to the rest-frame 
2--10~keV band was then converted into a count rate using the
exposures derived from the time-maps at each source position, 
and then into a flux assuming a power-law with photon index
$\Gamma=1.4$. To provide support to our results,
X-ray spectra were also extracted for all of the 23 sources and
fitted using {\sc XSPEC} \citep[version 12.6;][]{arnaud}
with a a power-law modified by Galactic absorption only. 
The derived rest-frame 2--10~keV fluxes were found to be consistent, 
within their admittedly large errors, with the fluxes obtained by the 
simple count rate to flux conversion described above.

For 46 of the remaining 48 \nev-selected sources\footnote{For two  sources, 
no reliable X-ray photometry could be obtained from the available data.} 
with no individual X-ray detection, 
we have derived upper limits to the X-ray flux using the same
procedure described above: at each source position, counts have been
extracted taking into account the PSF size and shape in the observed band
equivalent to the rest-frame 2-10 keV energy range. One-sigma count
upper limits have then been converted into count rates using the
time-maps, and these were finally 
converted into X-ray flux upper limits assuming a power-law with 
photon index $\Gamma=-0.4$, which is broadly consistent with a 
Compton-thick spectrum.

As a final remark, we note that the population traced by X-ray
detected \nev-selected sources represents a limited fraction 
of the overall, likely obscured, AGN population with an X-ray
counterpart and no broad emission line in the optical spectra
(23/180, i.e. 12.7\%). 
The X-ray selection allows us to collect a larger number of z$\approx$1~\tyii~AGN
than the \nev-detection technique (180 vs. 71). 
Both samples include a population of heavily obscured AGN, as suggested by
the distribution of column density for \tyii~AGN in the Chandra-COSMOS survey
\citep{Lanzu13} on the one hand and by the large fraction of 
X-ray undetected \nev-selected sources (48/71) on the other hand.
In other words, the selection based on the lack of broad optical
emission lines coupled to relatively strong ($>10^{42}$~erg~s$^{-1}$) 
X-ray emission is somehow complementary to the selection of AGN based
on the presence of \nev\ emission.
%
%
%
   \begin{figure}
   \resizebox{\hsize}{!}{\includegraphics{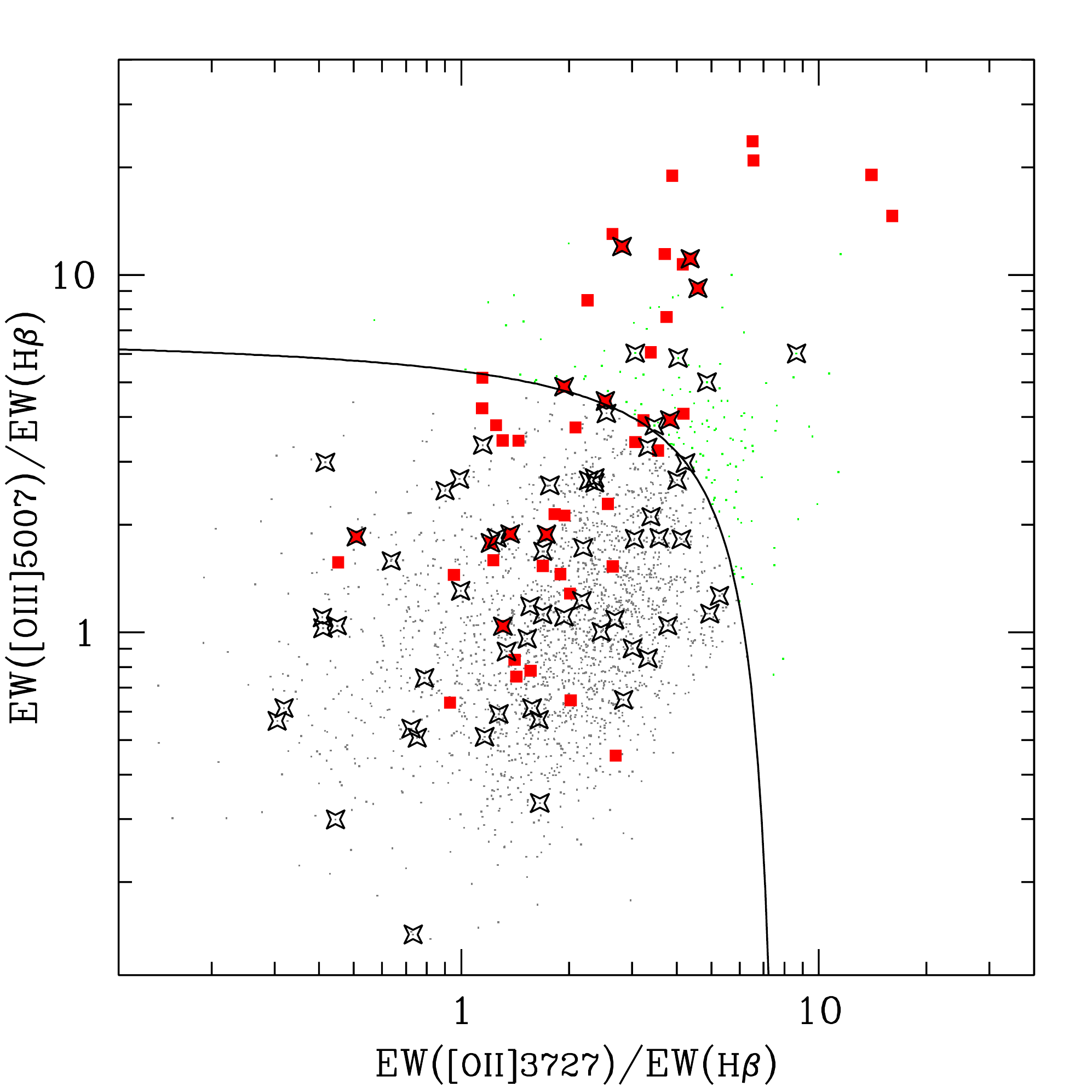}}
      \caption{``Blue" Diagnostic Diagram in the interval 0.65$\,<\,${\it z}$\,<\,$0.92. 
      The solid line shows the demarcation between SFG and \tyii~AGN
      defined by \citet{Lamare10}. Small points represent the 2461 zCOSMOS
      galaxies for which all the emission lines are detected with S/N$>$3.
      Large red squares correspond to \nev-selected objects, while
      the X-ray~emitting galaxies are identified by large starred symbols. 
         \label{FigblueDD}
      }
   \end{figure}

  \subsection{The ``blue'' line ratios diagram}

The classical method to classify star-forming galaxies and narrow line AGN
with optical spectroscopy is based on emission-line ratio diagnostics.
The most efficient and commonly used BPT diagram requires two sets
of line ratios: \nii~$\lambda$6583/\Halpha \ and \oiii~$\lambda$5007/\Hbeta.
Unfortunately, in the redshift range of interest here, the \Halpha \ line
is outside the observed spectral range, so
we need to rely on a diagnostic diagram based only on emission
lines observed in the blue part of the optical spectra: \oiii~$\lambda$5007,
\Hbeta \ and \oii$\lambda$$\lambda$3726,3729. Such ``blue" diagnostic
diagram, recently applied to a smaller (10k) sample of zCOSMOS galaxies
\citep{Bongiorno10}, is less effective than the lower redshift, classical BPT
diagnostic diagram.
The emission line fluxes of the zCOSMOS galaxies were
measured using the automated pipeline {\it platefit\_vimos} \citep{Lamare09},
which simultaneously fits all the emission lines with Gaussian functions
after removing the stellar continuum. We checked the consistency of the
platefit measurements with the \oii~fluxes measured by us in the sample
of \nev-selected AGN, finding an excellent agreement: the mean of the
intensity ratio distribution was 0.997 and two flux measurements never 
differed by more than 20\%.

%
%
%
%
The sample of zCOSMOS galaxies in the redshift range 0.65$\,<\,${\it z}$\,<\,$0.92
includes 5662 objects; this number includes 63 out of 94 \nev-selected \tyii~AGN.
The upper redshift limit is chosen to ensure that the \oiii~line is still
within the observed spectral range. 
The basic requirement for a reliable classification is that all the
diagnostic emission lines are detected above a minimum signal-to-noise
ratio (S/N). In particular, we selected emission-line galaxies in the explored
redshift range for which all the lines are detected with S/N$>$3.
This S/N cut reduces the number of analyzed galaxies down
to 2461 (43\%), with the strongest criterion being that on \Hbeta, which is generally
the weakest of the involved emission lines. A larger fraction (48/63; more than 75\%)
of \nev-selected \tyii~AGN satisfies the requirements for being included in
the diagnostic diagram shown in Figure~\ref{FigblueDD}. This diagram allows us to
separate the zCOSMOS star-forming galaxies from the Seyfert~2-like objects
which inhabit the upper part of the line ratio plane.  The demarcation line,
proposed by \citet{Lamare10}, separates the underlying galaxy population
in 159 \tyii~AGN (6.5\%) and 2302 SFG. The position of the \nev-selected
narrow line zCOSMOS galaxies is also shown (red square dots) in
Figure~\ref{FigblueDD}: a significant fraction (29/48 or 60\%)
of \tyii~AGN selected via \nev \ fall in the star-forming region of the
blue diagnostic diagram. Conversely, of the 159 \tyii~AGN selected
on the basis of their emission line ratios, only 19 (12\%) show a detectable
\nev \ emission. As a further check, we stacked the optical spectra of the
remaining 140 \tyii~AGN, selected by the diagnostic diagram, detecting 
a very faint \nev \ emission line with a rest-frame EW of 0.7\AA.
In Figure~\ref{FigblueDD} we also highlighted the emission-line galaxies with
an X-ray counterpart in the Chandra catalog (open star symbols):
again, 55 out of 66 (83\%) X-ray sources (with L$_{0.5-10 keV}>10^{42}$~erg~s$^{-1}$,
thus likely \tyii~AGN) fall in the region of star-forming galaxies. So, the
\nev-selected and X-ray~emitting \tyii~AGN share similar classification percentage
and position in the blue diagnostic diagram, suggesting a heavy incompleteness
in the line-ratios technique used to select high redshift  ($z > 0.5$) obscured AGN.

The lower efficiency in selecting AGN of the blue diagnostic diagram has
been already reported in literature \citep{Bongiorno10, Stasinska06},
with a number of possible explanations: first, because of the large wavelength
separation between the \oiii~$\lambda$5007 and \Hbeta \ emission lines,
their flux ratio is very sensitive to dust obscuration on galactic scale, and the
adoption of an EW~ratio can only mitigate the effect, due to the differential
extinction between the emission line region and the stellar continuum in
the galaxy spectra. 
Another possible reason for the presence of \tyii~AGN in the SFG region is
that in these objects star formation and AGN activity coexist. This
hypothesis is consistent with model predictions for the position of the
composite AGN/SF in the optical diagnostic diagrams \citep{Stasinska06}.
The latter explanation seems more plausible for the \nev~selected \tyii~AGN
falling in the SFG region, since a small amount of extinction in the host
galaxy would erase the blue faint \nev~emission line.

  \subsection{The Mass-Excitation (MEx) diagnostic}
The Mass-Excitation diagnostic diagram has been recently proposed by \citet{MEx}
and offers, at $z\ge0.5$, a more complete AGN selection than the
optical blue diagram. It is derived from the classic BPT diagram, using the
galaxy stellar mass as a surrogate for the \nii~$\lambda$6583/\Halpha \ line
ratio. \citet{MEx} demonstrate that the MEx technique successfully distinguishes
between star formation and AGN emission, also dealing with AGN/SF composite
galaxies, the so-called ``MEx-intermediate" class, which lie in a part of
the diagram in between the AGN and SFG regions.
%
   \begin{figure}
   \resizebox{\hsize}{!}{\includegraphics{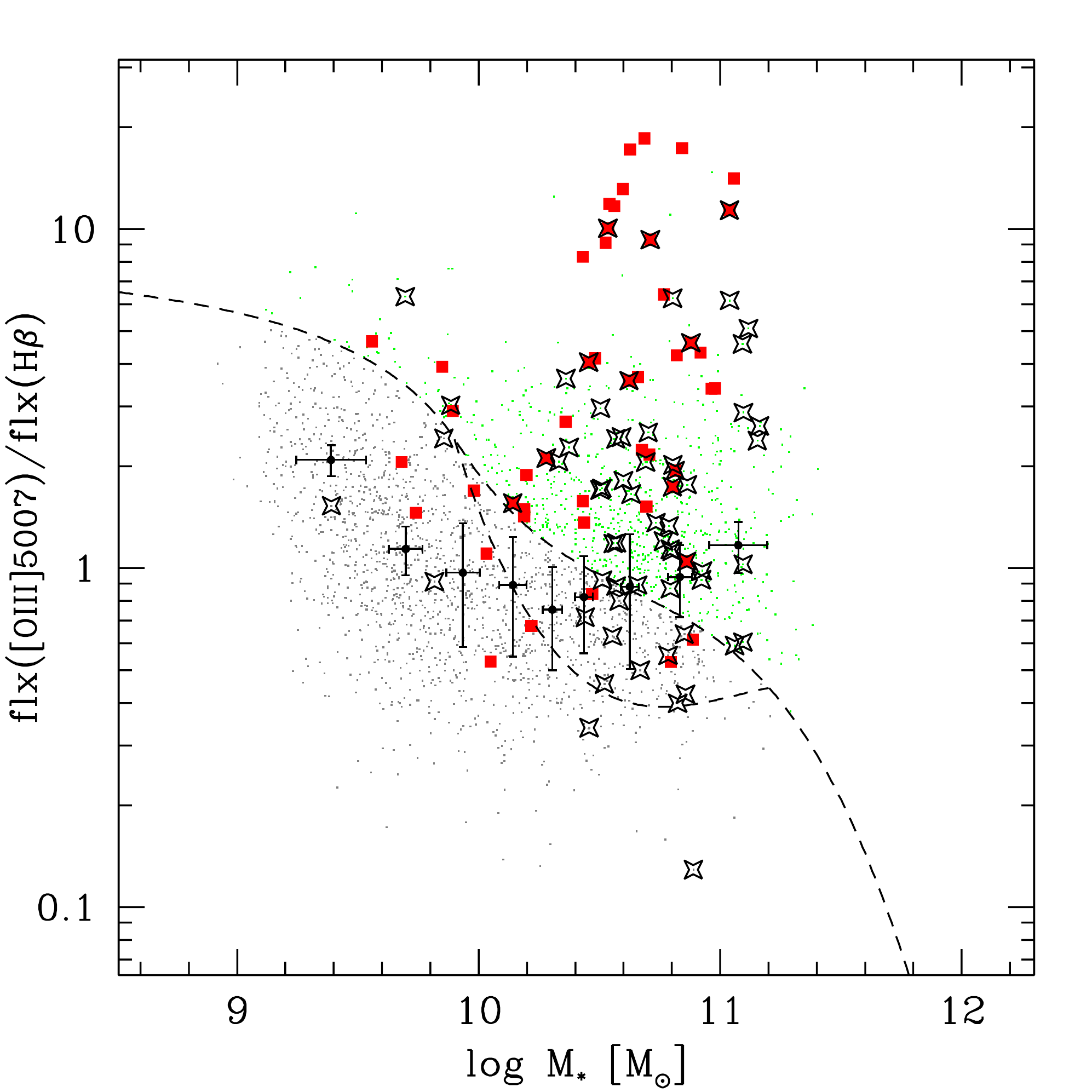}}
      \caption{The Mass-Excitation (MEx) diagram in the redshift range 0.65$\,<\,${\it z}$\,<\,$0.92. 
      The dashed lines show the empirical curves defined by \citet{MEx},
      dividing the plane in an upper region, occupied by galaxies hosting an AGN,
      and in a lower region where the star-forming galaxies are located.
      Objects located between the two curves are classified as AGN/SF composite
      galaxies. The small points represent the same sample of 2461 zCOSMOS
      galaxies used in Figure~\ref{FigblueDD}.
      Large red squares correspond to \nev-selected objects, while
      the X-ray~emitting galaxies are identified by large starred symbols. 
      The filled dots with error bars represent the line flux ratios measured in nine
      composite spectra of galaxies that individually were not included in the
      diagram because of their low S/N emission lines. The horizontal error bars
      indicate the dispersion of the stellar mass values in each of the 
      bin, whereas the vertical error bars show the estimated error on the flux ratios.
         \label{FigMExDD}
      }
   \end{figure}

%
%
The same zCOSMOS galaxy sample adopted for the blue diagnostic diagram
(objects in the range 0.65$\,<\,${\it z}$\,<\,$0.92 and with emission lines all
detected with S/N$>$3) has been analyzed with the MEx diagnostic diagram
presented in Figure~\ref{FigMExDD}.
The high efficiency in selecting AGN of the MEx technique is evident when
looking at the position of the \nev-selected (red squares) and X-ray~emitting
(starred symbols) \tyii~AGN. A large fraction (39/48, more than 80\%) of the
\nev-selected galaxies are classified as AGN, along with other four objects lying
in the region of the diagram where AGN/SF composites are expected. Only
five of them would be classified as star-forming galaxies by the MEx diagnostic
diagram. Similarly, of the 66 X-ray~emitting galaxies which satisfied the 
constrain for the inclusion in the diagram, fifty are classified as AGN and
eleven belong to the MEx-intermediate class. Once again, although the
two samples have only few objects in common, the \nev-selected
and X-ray~emitting \tyii~AGN share almost identical position and classes
percentage in the MEx diagnostic diagram.

However, a few words of caution about the efficiency of the MEx technique
are appropriate. The inclusion of a very large percentage of AGN selected
with other techniques is probably the consequence of a quite ``generous"
criterion for AGN classification by the MEx diagnostic. When it is applied
to the zCOSMOS galaxies in the range 0.65$\,<\,${\it z}$\,<\,$0.92, about
one-third of them (796) are classified as active, a quite large number of AGN
if compared to those identified by other techniques. Moreover,
since more than 3000 galaxies are not included in the MEx diagram because of 
low S/N in their emission line measurements, we had to estimate how their
removal would bias the AGN fraction. We divided the excluded galaxies
in nine equally populated bins according to their computed stellar mass, and
fluxes of the emission lines have been measured in composite spectra. 
In seven over nine bins the value of \oiii/\Hbeta \ flux ratio, plotted in
Figure~\ref{FigMExDD} against the mean stellar mass, falls below the MEx
empirical division curve, suggesting that zCOSMOS galaxies with low S/N
spectra predominantly populate the star-forming region, and, if they were
included in the diagnostic diagram, the AGN percentage would decrease to
$\approx 25$\%. In addition, since the empirical dividing lines used in 
Figure~\ref{FigMExDD} have been calibrated in \citet{MEx} using a {\it z}$\sim$0.1
SDSS galaxy sample, their application to our {\it z}$\sim$0.8 sample 
should be taken with caution. Indeed, new computations carried out by Juneau et~al.
(in prep.) confirm that the loci of these diagnostic diagrams evolve with redshift,
and that approximately 10\% of the galaxies falling in the AGN class using
local relation are instead included in the SF region at {\it z}=0.7. This would
further decrease, by a similar amount, the MEx-AGN fraction in our zCOSMOS
galaxy sample.
 
  \subsection{The relative efficiency of  \tyii~AGN selection techniques}
%
   \begin{table}
      \caption[]{Numbers of zCOSMOS emission-line galaxies classified as \tyii~AGN by different selection techniques.}
         \label{seleffty2}
     $
       \resizebox{0.485\textwidth}{!}{
	\begin{tabular}{lcccc}
            \hline
            \noalign{\smallskip}
            & \nev & X-ray & DD-Blue  & DD-MEx \\
            \hline
            \noalign{\smallskip}
            \nev & {\bf 36 (2.1}\%) & 11 (17\%) & ~12 (10\%) & ~30 (5.1\%) \\
             \noalign{\smallskip}
           X-rays & 11 (31\%) & {\bf 66 (3.9}\%) & ~11 (9.4\%) & ~50 (8.6\%) \\
            \noalign{\smallskip}
            DD-Blue & 12 (33\%) & 11 (17\%) & {\bf 117 (6.8}\%) & ~69 (12\%)\\
            \noalign{\smallskip}
            DD-MEx & 30 (83\%) & 50 (76\%) & ~69 (59\%) & {\bf 582 (34}\%) \\
            \noalign{\smallskip}
            \hline
         \end{tabular}
       }
     $
  \tablefoot{The diagonal elements represent the number of galaxies classified as
    \tyii~AGN by each technique, with corresponding efficiencies respect to the 
    sample of 1712 galaxies for which all diagnostics are available. \tyii~AGN 
    identified by two methods are shown as off-diagonal elements, along with 
    the percentage with respect to the total number of galaxies (boldface number in
    same column) selected by the specific diagnostic identified in the column top label. 
   }
   \end{table}
In this section the previously discussed optical and X-ray diagnostics are
compared. 
Since the different selection techniques are not applicable to all the members
of the parent galaxy sample, in order to obtain a
meaningful comparison, we extract a subsample of 1712 galaxies (included
in the Chandra field-of-view, with a redshift range of 0.65$\,<\,${\it z}$\,<\,$0.92,
and with emission lines all detected with S/N$>$3) for which all diagnostics are available.
The objects that show at least one indication of activity are 651, more than one-third
of the galaxy population (38\%) but, as already demonstrated in Section~5.3,
the exclusion of galaxies with low S/N emission lines in the MEx diagram may
significantly increase the AGN percentage. If we do not take into
account the MEx diagnostic, the AGN number drops to 191 (11\%).

A relative evaluation of the different techniques, presented in this paper,
to select \tyii~AGN at z$\approx$1 can be drawn from Table~\ref{seleffty2}:  
the first three selections (namely, \nev \ and X-ray detections, and the
blue line ratios diagram) are almost complementary, with similar efficiency
and relatively small overlaps, since the fraction of objects simultaneously
detected by two methods ranges from 10 to 33\%.
Conversely, the MEx diagnostic seems highly efficient in selecting obscured active
galaxies, since more than 50\% of the \tyii~AGN\footnote{Taking in account also
the redshift evolution of the empirical dividing lines.} identified by other methods are
located in the AGN region of the MEx diagram.
The differences in the nature of the selected \tyii~AGN samples
play an important role in the relative merits of the selection methods investigated.
The \nev- and X-ray-selected galaxies have clear sign of nuclear activity,
representing good single examples of powerful AGN. On the other hand, the
MEx selection, with its probabilistic approach, produces a 
more complete {\it statistical} AGN sample, including also low-luminosity
objects, but it can be affected by contamination especially when used
on an object-by-object basis.

\section{The X-ray to \nev \ flux ratio and the heavily obscured AGN content
in zCOSMOS}

\citet{Gilli10} explored the potential of the observed \hbox{2--10 keV} 
to \nev\ emission-line luminosity ratio (X/NeV) as a method to
discover heavily obscured, possibly Compton-thick AGN up to z$\sim$1.5.
The zCOSMOS \nev-selected \tyii~AGN provide an ideal sample to 
apply this promising diagnostic.
%
\begin{figure}
\resizebox{\hsize}{!}{\includegraphics{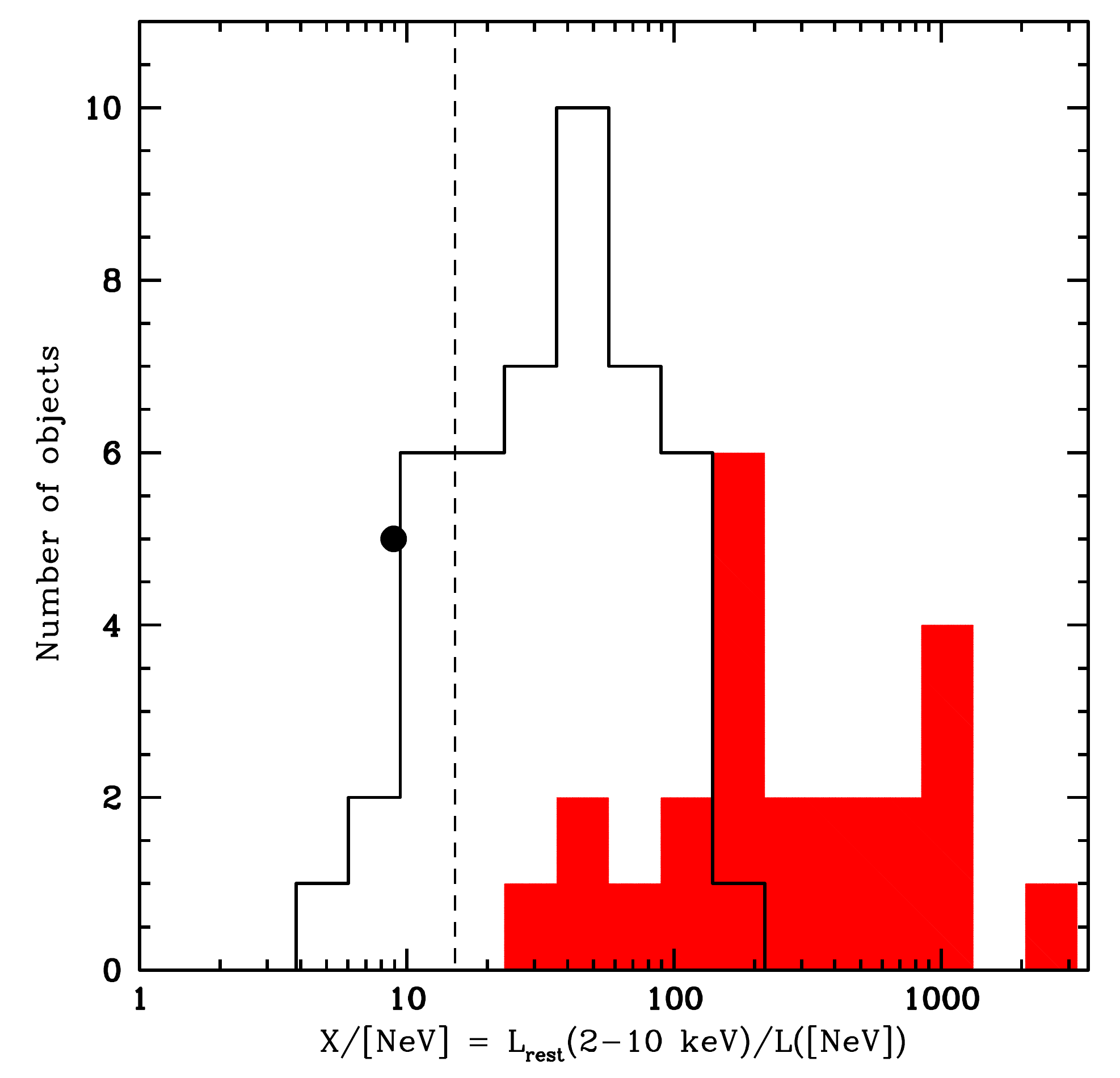}}
\caption{The distribution of the observed (i.e., not corrected for 
absorption/extinction) rest-frame \hbox{2--10 keV} to \nev \ luminosity ratio
(X/NeV) for the 69 zCOSMOS \tyii~AGN with reliable X-ray photometry 
(23 X-ray detections and 46 upper limits, see Section 5.1 for details). 
The filled histogram represents the 
X-ray detected objects, while the histogram delimited by the solid
line shows the luminosity ratio upper limits for  X-ray undetected \tyii~AGN. 
The vertical dashed line indicates the threshold defined by 
\citet{Gilli10}: objects to the left of this line are candidate
Compton-thick AGN. 
The filled circle indicates the average value 
obtained from stacking the sample of 46 X-ray undetected \tyii~AGN.
\label{FigX_NeV}
}
\end{figure}

In order to compare the \nev \ line flux measured in zCOSMOS spectra
with the integrated Chandra X-ray emission in a fair manner, we need to
correct the optical spectral measurements for slit losses.
In the redshift interval of interest (z$\ge$0.65), the 1\arcsec-slit width used in zCOSMOS
observations corresponds to a physical size greater than 7~kpc, so, unless the sizes
of the NLR are unreasonably larger than the typical value of $\approx$1~kpc, the line
flux that enters the spectrograph aperture can be considered as coming
from a point source, and it is essentially affected by seeing variations only. 
Thus, we corrected the \nev \ flux for each of our objects using
aperture corrections estimated through spectroscopic stars 
observed in the same multi-slit mask of the analyzed AGN spectrum. 
In our \tyii~AGN sample, the aperture corrections range between 0.03 and 1.69 mag,
with a median value of 0.46 mag. Ten of the \nev-selected galaxies had multiple
zCOSMOS spectra, and applying the estimated aperture correction allowed us to
lower, on average, the relative flux differences of the repeated measurements
from 17\% to 9\%. The X-ray fluxes are computed in the \hbox{2--10 keV} rest frame
band, without any absorption correction, in order to follow the definition given in \citet{Gilli10}.

The distribution of the X/NeV ratio for the 69 zCOSMOS \tyii~AGN with
reliable Chandra measurements (see Section 5.1 for details) 
is plotted in Figure~\ref{FigX_NeV}. \citet{Gilli10} showed that, in a
sample of local Seyferts, the AGN with X/NeV$<$15 are almost all Compton-thick. 
In our \nev-selected sample, 9 \tyii~AGN are below this threshold,
being all undetected in the X-rays. These sources can be safely considered
robust Compton-thick AGN candidates. 
Of the the 46 \nev-selected \tyii~AGN with X/NeV upper limits,
39 have X/NeV$<$100, suggesting a column density possibly exceeding 
$10^{23}$ cm$^{-2}$ \citep{Gilli10}. 
Conversely, all of the 23 X-ray detected objects fall in the region
likely populated mostly by Compton-thin AGN (with 
30$\,<\,$X/NeV$\,<\,$2100). 

Synthesis models of the X-ray background \citep{TU06, GCH07}
predict that the ``missing" XRB at 30~keV (i.e. the XRB emission that
cannot be accounted for by unobscured and moderately obscured AGN)
is indeed produced by Compton-thick objects. 
According to the synthesis model of \citet{GCH07}, most of the missing 
XRB is produced by Compton-thick AGN in the redshift interval \hbox{z=0.5--1}
and with intrinsic \hbox{2--10 keV} luminosities
in the range $10^{42-44}$~erg~s$^{-1}$: these intervals match
almost perfectly with those of our \nev-selected
sample\footnote{The \nev~or \oiii \ luminosities are converted
into \hbox{2--10 keV} intrinsic luminosities following the recipes
described in \citet{Vignali10} and \citet{Gilli10}}.

Our sample is therefore placed in the best position to assess the
relevance of Compton-thick AGN to the XRB emission. 
To this purpose, we used X-ray stacking analysis to derive the average
X/NeV ratio for the 46 X-ray undetected \nev-selected \tyii~AGN
(represented by the empty histogram in Figure~\ref{FigX_NeV}) and for two
sub-samples of these X-ray undetected sources, those with X/NeV$<$30
and those with X/NeV$>$30. The ``dividing line'' of X/NeV=30 has been
chosen to have a similar number of objects in both sub-samples (22 and
24, respectively). 
For the 46 X-ray undetected \nev-selected \tyii~AGN, the derived
average X/NeV ratio is 9.8 (filled circle in Figure~\ref{FigX_NeV}), 
while for the X/NeV$<$30 (X/NeV$>$30) sub-sample the ratio is 3.6 (36).
We note that for the first sub-sample X-ray photometry in the
rest-frame \hbox{2--10~keV} band provides only a 1.6$\sigma$ detection, while
for the second sub-sample we have a 5.8$\sigma$ detection. 
The results of the stacking analysis were used to tighten the constraints on the 
Compton-thick AGN fraction among the X-ray undetected sources as follows:
an average of 2.6, 1.1 and 3.9 net counts per source were obtained 
when stacking the total sample of 46 X-ray undetected sources, 
the ``faint'' sub-sample of 22 sources with X/NeV$<$30, 
and the ``bright'' sub-sample of 24 sources with X/NeV$>$30, respectively. 
We then assumed that the count statistics of the 46 X-ray undetected 
sources follows a Poisson distribution with mean 2.6, and verified 
that this is a good assumption since the integration of the lower 
and upper half of such a distribution returns an average counts
of 1.2 and 3.9, which are in excellent agreement with the values
measured for the ``faint'' and ``bright'' sub-samples. 
We then performed 10$^4$ Montecarlo runs, each time randomly extracting 46 
count values from the Poisson distribution and converting these counts 
into X-ray fluxes with the prescriptions given in Section 5.1. These random
X-ray flux catalogs were associated to the NeV flux catalog of X-ray undetected 
sources and the corresponding X/NeV ratios were then computed. 
By considering the whole set of simulations, on average 29.4 sources 
(with 2.7 rms) were found to have X/NeV$<$15 and are therefore Compton-thick candidates. 
Based on the above simulations, we estimate that the Compton-thick 
fraction in our sample of 69 \tyii~AGN is $43\pm4\%$, 
which is broadly consistent with the expectations from XRB synthesis models 
(e.g. 50\% in \citealt{GCH07}). 
We stress that the small uncertainties in our estimate of 
the Compton-thick AGN fraction are just statistical, 
whereas the systematics related to the selection method and stacking analysis 
are likely to dominate the error budget. 
Significant improvements are likely to be obtained
by using the data from the recently
approved Chandra COSMOS Legacy Survey (2.8 Ms, PI F. Civano), which
will observe the outer portion of the COSMOS field and enlarge the
total X-ray coverage to 1.7~deg$^2$ at a depth of 160 ks: the 24
objects \nev-selected objects which are now outside the Chandra mosaic
should be then observed, and about half of the current X/NeV upper limits 
should be improved by a factor of 2 (or turn into real measurements)
by these new X-ray data. A deeper investigation of the Compton-thick
candidates among \nev-selected obscured AGN in COSMOS will be presented
in a forthcoming paper (Vignali et al., in preparation), by exploiting both
the X-ray techniques (stacking of undetected objects, spectral analysis)
and infrared data.

\section{Summary}

In this paper we have presented a method to select $z\sim\,$1 obscured AGN
from optical spectroscopic surveys, useful to gather a more complete
census of actively accreting black holes in galaxies. The detection of the
high-ionization \nev~$\lambda3426$ line has been used to pinpoint active
nuclei in the  20k-Bright zCOSMOS galaxy sample and we have successfully
found systems that escaped other AGN selection techniques.

%
%
   The \nev-selected \tyii~AGN sample consists of 94 sources in the
   redshift range 0.65$\,<\,${\it z}$\,<\,$1.20, spanning the \oiii \ 
   luminosity range $10^{7.5}\,$L$_{\sun}\,<\,$L\oiii$\,<\,10^{9.0}\,$L$_{\sun}$.
   The \tyii~AGN composite spectrum closely resembles the spectrum of a
   Seyfert 2 galaxy with strong high-ionization narrow lines. Nevertheless,
   the line ratios would place it in the transition region of diagnostic diagrams,
   suggesting some star formation contamination.
   The mean optical extinction of the narrow line AGN sample, estimated from
   the composite \Hbeta/\Hgamma\ flux ratio, is \hbox{$\langle$E(B$-$V)$\rangle\,$=$\,0.18$}. 
   The absorption-line continuum of the composite spectrum, fitted with
   BC03 population synthesis models, gives information on the average stellar
   content of the hosts, that roughly corresponds to an Sa/Sb galaxy.
  
   In the same redshift interval, the zCOSMOS survey discovered 60 broad-line
   AGN, 37 of them with a detected \nev~$\lambda3426$ line. Comparing the
   composite spectra of both the full \tyi~sample and the \nev-detected
   \tyi~sub-sample, no significant differences have been found, which indicates that
   the latter does not show special optical characteristics with respect to the full 
   sample. The non-detection of the faint [Ne v] emission line in a fraction
   of \tyi~AGN spectra is therefore probably due to the presence of a stronger
   continuum in these objects, continuum that is shielded in the NL-sample.
   The two composite spectra do not show peculiar properties also with respect 
   to other known quasar sample.
   
   The main emission lines have also been accurately measured in all
   the zCOSMOS spectra of \nev-selected objects. The average \oii/\nev \
   flux ratio in the \tyii~AGN sample ($\approx$4) is significantly larger than in
   the \tyi~AGN sample ($\approx$2). If interpreted as due to an excess in the
   \oii \ luminosity, the larger line ratio is suggestive of enhanced star formation
   in the obscured population, at least with respect to the sample of \tyi~AGN.

   The stellar masses of galaxy hosts 
   cover the range $5\times 10^{9}\,$--$\,3\times 10^{11} M_\odot$, and are,
   on average, higher than those of the galaxy parent sample. The median
   stellar mass of the \tyii~AGN hosts is $8\times 10^{10}M_\odot$, while
   for the parent galaxy sample is $3.7\times 10^{10}M_\odot$. 
   The fraction of galaxies hosting \nev-selected \tyii~AGN
   increases with the stellar mass, reaching a maximum of around 3\% at
   $\approx$2$\,\times\,10^{11}M_\odot$. 

   A visual morphological classification has been assigned to
   71 out of 94 (76\%) of the \tyii~AGN host galaxies using
   the ACS images available in COSMOS.
   Comparing the AGN host morphologies with those of a
   carefully mass-matched sample of normal galaxies, the two distributions
   do not look strikingly different, although a possible trend in the
   relative frequency of the Hubble types seems to emerge:
   the [Ne v]-selected Type-2 AGN do prefer the
   early-spirals (Sa-Sb) galaxies, with an expected lack of
   late-spirals and Irregulars morphologies. Moreover, a lower fraction
   of elliptical galaxy hosts, at least with respect to the control sample,
   is observed. The host galaxy population, which shows later morphologies
   with respect to brighter \tyii~AGN samples
   \citep[i.e. the SDSS sample of][who found mainly elliptical hosts]{Zakamska06},
   is probably related to the different nuclear luminosity of the samples:
   our \nev-selected objects cover an \oiii \ luminosity range at least
   an order of magnitude fainter than SDDS \tyii~quasars.

   The selection techniques of \tyii~AGN at z$\sim$1 have also been 
   investigated, with respect to their relative efficiency. First, the \nev-selected
   sample has been compared with the \hbox{C-COSMOS} catalog of X-ray
   sources: the \nev \ technique discovered a large fraction of \tyii~AGN
   (46/69, 67\%) undetected in the medium-depth, wide-area Chandra survey.
   Many of these X-ray faint, \nev~emitting galaxies may be heavily
   obscured active nuclei, especially if we consider that, of the 39 zCOSMOS broad
   line AGN falling in the Chandra region, all but two are detected in the X-rays.
   Conversely, a substantial number of X-ray luminous galaxies (158) do not
   show a detectable \nev \ emission, neither in single zCOSMOS spectra,
   nor in the stacked optical spectrum.  
   Similar results have been obtained in comparing \nev \ and line ratio selection
   methods: the blue diagnostic diagram, based on \oiii/\Hbeta \ versus \oii/\Hbeta \ line
   ratios, includes 48 \nev-selected \tyii~AGN, and again a large fraction of them 
   (29/48, 60\%) fall in the star-forming region of the diagram.
   Likewise, 55 out of 66 (83\%) luminous X-ray sources
   fall in the region of star-forming galaxies. The \nev-selected and X-ray-emitting
   \tyii~AGN share similar position and class distribution in the
   \oiii/\Hbeta \ versus \oii/\Hbeta \ plane, but these two methods alone cannot
   provide a fairly complete \tyii~AGN selection, since a significant number of
   emission line galaxies (136), without a visible \nev~$\lambda3426$ line in the
   spectra and undetected in X-rays, are classified as active by the diagnostic diagram.
   Finally, the Mass-Excitation diagnostic diagram seems the most comprehensive
   of the analyzed selection techniques, since more than 50\% of the \tyii~AGN identified
   by other methods are classified as AGN according to the MEx diagram. However,
   MEx-AGN classification may not always hold on an individual galaxy basis and
   the degree of contamination is probably the main drawback of the method,
   as only $\approx$20\% of the MEx-AGN candidates are selected by one of the other
   techniques.
  
   Finally, the \hbox{2--10 keV} to \nev\ emission line luminosity ratio has been 
   exploited to search for the more heavily obscured AGN.
   For the 69 \nev-selected \tyii~AGN with reliable Chandra measurements, 
   a significant fraction (46 objects, 67\%) is undetected in the X-rays and
   only X/NeV upper limits can be derived. We exploited X-ray stacking analysis
   to estimate the average X/NeV ratio for the X-ray undetected \nev-selected
   galaxies. A set of Montecarlo simulations, based on simple assumptions,
   was performed, finding that the Compton-thick fraction in our sample
   of \tyii~AGN is of the order of 40\%, in good agreement with the XRB
   synthesis models \citep{GCH07}.
   Conversely, all the 23 X-ray detected, \nev-selected \tyii~AGN fall in the
   region of the Compton-thin objects. We intend to further investigate the
   nature of these Compton-thick AGN candidates in a forthcoming paper.

\begin{acknowledgements}
      This work was supported by the INAF Grant ``PRIN--2010". 
We also acknowledge financial contribution from the agreement ASI-INAF I/009/10/0.
and from the ``PRIN--INAF 2011". The authors would like to thank the referee for
his/her valuable suggestions and St\'ephanie Juneau for helpful discussions
and for sharing her results prior to publication.
\end{acknowledgements}

\end{document}